\documentclass[%
reprint,
pra,
]{revtex4-2}

\usepackage[utf8x]{inputenc}
\usepackage{graphicx}

\usepackage{amsmath}
\usepackage{amssymb}
\usepackage{siunitx}
\usepackage{mathtools}

\usepackage{hyperref}
\usepackage[pdftex,dvipsnames]{xcolor}
\hypersetup{frenchlinks = true,
	colorlinks,
	linkcolor={red!50!black},
	citecolor={blue!50!black},
	urlcolor={blue}
}

\usepackage{lipsum}
\usepackage{braket}
\usepackage{enumitem}

\usepackage{lmodern}
\usepackage{slantsc}

\usepackage{xargs}                      
\usepackage[colorinlistoftodos,prependcaption,]{todonotes}
\newcommandx{\unsure}[2][1=]{\todo[linecolor=red,backgroundcolor=red!25,bordercolor=red,#1]{#2}}
\newcommandx{\change}[2][1=]{\todo[linecolor=blue,backgroundcolor=blue!25,bordercolor=blue,#1]{#2}}
\newcommandx{\info}[2][1=]{\todo[linecolor=OliveGreen,backgroundcolor=OliveGreen!25,bordercolor=OliveGreen,#1]{#2}}
\newcommandx{\improvement}[2][1=]{\todo[linecolor=Plum,backgroundcolor=Plum!25,bordercolor=Plum,#1]{#2}}
\newcommandx{\thiswillnotshow}[2][1=]{\todo[disable,#1]{#2}}

\renewcommand{\Re}{\mathrm{Re}}

\paperwidth=\dimexpr \paperwidth + 6cm\relax 
\oddsidemargin=\dimexpr\oddsidemargin + 3cm\relax
\evensidemargin=\dimexpr\evensidemargin + 3cm\relax
\marginparwidth=\dimexpr \marginparwidth + 3cm\relax

\def\mat#1{\mathbf{#1}}
\def\op#1{\hat{#1}}
\def\st#1{_{\mathrm{#1}}}
\def\ut#1{^{\mathrm{#1}}}

\renewcommand{\d}[0]{\mathrm{d}}

\newcommand{\U}[0]{\op{\mathcal{U}}}
\renewcommand{\H}[0]{\op{H}}

\newcommand{\Hd}[0]{\H^{d}}
\newcommand{\Hc}[0]{\H^{c}}
\newcommand{\R}[0]{\op{\mathcal{R}}}
\newcommand{\psio}[0]{\psi\st{ini}}
\newcommand{\dt}[0]{\delta t}
\newcommand{\F}[0]{F}
\newcommand{\overlap}[0]{o}
\newcommand{\nt}[0]{N_{t}}

\usepackage{xspace}
\newcommand{\grape}{\textsc{grape}\xspace}

\newcommand{\psitgt}[1][]{%
	\ifthenelse{\equal{#1}{}}{\psi_{\mathrm{tgt}}}{\psi_{\mathrm{tgt,#1}}}%
}

\newcommand*\olinepossiblility[1]{%
	\vbox{%
		\hrule height 0.5pt
		\kern0.25ex
		\hbox{%
			\kern-0.25em
			\ifmmode#1\else\ensuremath{#1}\fi
			\kern-0.05em
		}
	}
}

\newcommand*\oline[1]{\overline{#1}}

\usepackage{bm}
\renewcommand\vec{\bm}
\def\mat#1{\vec{#1}} 

\begin{document}
	\title{Approximate Dynamics Lead to More Optimal Control: Efficient Exact Derivatives}
	\author{Jesper Hasseriis Mohr Jensen$^1$}
	\email{jhasseriis@phys.au.dk}
	\author{Frederik Skovbo Møller$^{1,2}$}
	\author{Jens Jakob Sørensen$^1$}
	\author{Jacob Friis Sherson$^1$}
	\email{sherson@phys.au.dk}
		\affiliation{%
		$^1$ Department of Physics and Astronomy, Aarhus University, Ny Munkegade 120, 8000 Aarhus C, Denmark
	}
	\affiliation{%
		$^2$ Vienna Center for Quantum Science and Technology, Atominstitut, TU Wien, Stadionallee 2, 1020 Vienna, Austria
	}

	\begin{abstract}
	Accurate derivatives are important for efficiently locally traversing and converging in quantum optimization landscapes. 
	By deriving analytically exact control derivatives (gradient and Hessian)
	for unitary control tasks, we show here that the computational feasibility of meeting this accuracy requirement depends on the choice of propagation scheme and problem representation.
	Even when exact propagation is sufficiently cheap it is, perhaps surprisingly, much more efficient to optimize the (appropriately) approximate propagators: approximations in the dynamics are traded off for significant complexity reductions in the exact derivative calculations. 
	Importantly, past the initial analytical considerations, only standard numerical techniques are explicitly required with straightforward application to realistic systems.
	These results are numerically verified for two concrete problems of increasing Hilbert space dimensionality. The best schemes obtain unit fidelity to machine precision whereas the results for other schemes are separated consistently by orders of magnitude in computation time and in worst case 10 orders of magnitude in achievable fidelity. 
	Since these gaps continually increase with system size and complexity, this methodology allows numerically efficient optimization of 
	very high-dimensional dynamics, e.g. in many-body contexts, operating in the high-fidelity regime which will be published separately.
	\end{abstract}

	\maketitle


\section{Introduction}
\label{sec:introduction}
The demand for precise and fast quantum control extending into high-fidelity regimes places increasing emphasis on the role of optimization methodologies and their performance capacities. 
Identification and extraction of quantum \textit{optimal controls} have enjoyed theoretical and experimental success in numerous research areas \cite{glaser2015training}, such as superconducting qubits \cite{motzoi2009simple,egger2013optimized,goerz2017charting,montangero2018introduction,mogens2020global},
nuclear magnetic resonance systems \cite{kehlet2004improving,khaneja2005optimal,nielsen2007optimal,kallies2018cooperative,sorensen2020optimization}, 
nitrogen vacancy centers \cite{scheuer2014precise,dolde2014high,waldherr2014quantum,chou2015optimal}, 
cold molecules \cite{koch2004stabilization,koch2006making,de2011optimal,tibbetts2013optimal}, and
cold atoms \cite{doria2011optimal,van2016optimal,mundt2009optimalcontrol, jager2013optimal,cui2017optimal,patsch2018fast, larrouy2020fast},
to name a few.
\begin{figure}[t]
	\includegraphics[trim={0 0 0 0.2cm},clip,]{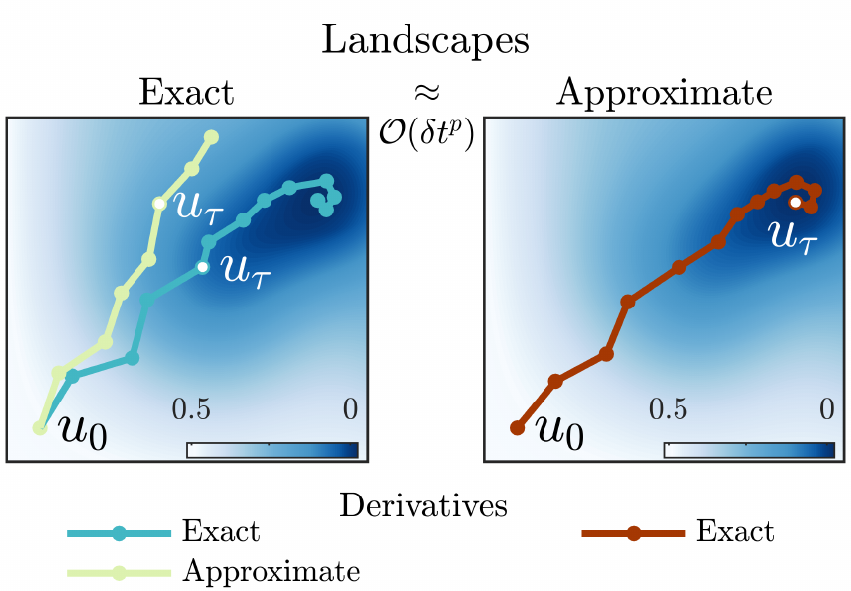}
	\caption{
		Abstract illustration of locally minimizing in exact and approximate landscapes (equivalent to dynamical schemes), each with different analytical forms for the exact derivatives.  
		The planes are spanned by control functions $u(t)$ with associated functional values given by the colormap and dots denote optimization iterates. 
		\textbf{Exact landscape:} exact derivatives are expensive but ultimately lead to optimal results, whilst cheaper, approximate derivatives may not. 
		\textbf{Approximate landscape:} exact derivatives and the dynamics itself are significantly cheaper and yield faithful optimal results under appropriate conditions. 
		Thus, for the \textit{same} optimization wall-clock time $\tau$, a common initial $u_0$ leads to different $u\st{\tau}$ 
		(white dot centers) and hence manifests a performance gap.
	}
	\label{fig:landscape}
\vspace{-0.5cm}
\end{figure}
At the same time, an increasing array of algorithmic approaches are available in these arenas, counting among others 
derivative-based (\grape \cite{khaneja2005optimal,de2011second,machnes2011comparing}, 
auxiliary matrix \cite{floether2012robust,goodwin2015auxiliary,goodwin2016modified} or equivalently \textsc{goat} \cite{machnes2018tunable}, \textsc{group} \cite{sorensen2018quantum,sorensen2020optimization}, Krotov \cite{tannor1992control,palao2002quantum,schirmer2011efficient}), derivative-free  (Nelder-Mead \textsc{crab} \cite{caneva2011chopped,doria2011optimal,van2016optimal}, stochastic ascent \cite{sels2018stochastic}, genetic evolutionary \cite{li2018global}), and combinations thereof \cite{sorensen2018approaching}.
Along a separate axis lies additional choices of open-loop \cite{dong2010quantum,caruso2012coherent}, closed-loop \cite{walmsley2003quantum,rosi2013fast,feng2018gradient}, and/or human-in-the-loop \cite{heck2018remoteopt,jensen2020crowdsourcing} control.

Irrespective of the physical platform and choice of optimization algorithm, a common denominator is inevitable: 
with growing problem complexity and numerical simulation efforts, the relative efficiency of each optimization cycle must be streamlined to allow convergence to e.g. high-fidelity solutions within finite time.
Accuracy and computational speed have been identified as important goals and challenges for modern control design \cite{glaser2015training,acin2018quantum}. In the context of derivative-based methods, i.e. update rules relying on local gradient and Hessian calculations of the optimization objective, this has been recognized at least since the seminal work presented in Ref.~\cite{khaneja2005optimal} where the analytical first-order approximation to the gradient was calculated.
However, this first-order approximation is not suited for obtaining standard quasi-Newton search directions due to the
rapid error accumulation in the Hessian approximation which is built iteratively from gradients \cite{nocedal2006numerical}.
The steepest descent direction is also only a minimally viable choice with the weakest convergence properties among the standard methods.
The use of quasi-Newton methods with the more desirable convergence properties was enabled later by e.g. Ref.~\cite{de2011second} where the analytically exact gradient for an exact propagator was calculated at the expense of additional computational time per iteration. 
As system sizes increase, however, exact propagators and their exact derivatives become prohibitively resource intensive. 

In this work, we advance the theoretical toolbox for obtaining controls that satisfy high-performance criteria in arbitrary unitary quantum tasks. 
Following a discretize-then-optimize approach we derive general \textit{analytically exact} gradients and Hessians for different propagation schemes, specifically an exact exponentiation propagator and two Suzuki-Trotter propagators which we interpret in terms of optimization landscapes. 
This means that each choice of effective time evolution operator gives rise to its own, but not necessarily dynamically exact, optimization landscape as illustrated in Fig.~\ref{fig:landscape} (Sec.~\ref{sec:exactderivatives}). 
Given, then, that the exact propagator approach is sometimes computationally infeasible, 
we thus examine the interplay between approximations in the landscape (i.e. dynamics) versus in the derivative calculations.

We show that the complexity of exact analytical derivatives strongly depends on the chosen propagation scheme, corresponding to the specifics of the numerical implementation details, and representation of the problem: solving the problem in a basis where the controllable part of the Hamiltonian is diagonal and simultaneously employing one of the Trotterized propagators greatly simplifies the derivative calculations (Sec.~\ref{sec:analytical}). 
Analytically exact derivatives can thus be computed very efficiently, principally limited only by the time it takes to propagate states which by virtue of the dynamical approximation is also particularly cheap.
That is, our results and e.g. Fig.~\ref{fig:landscape} are not just trivial consequences of the reduced propagation time due to the dynamical approximation, but also the complexity reduction of the analytical exact derivatives. 
Through numerical experiments, remarks on implementation details, scaling comparisons, and generalizability analyses, our main goal is to show that the presented Trotter derivative methodologies 
are not only very efficient on realistic problems, but also straightforwardly applicable since they rely only on otherwise well-known ingredients.
Another state-of-the-art approach to calculating exact gradients and Hessians is through the aforementioned so-called \textit{auxiliary matrix method} \cite{floether2012robust,goodwin2015auxiliary,goodwin2016modified} as implemented e.g. in the Spinach software library \cite{hogben2011spinach}, and we include this methodology in our comparative studies. 

We cement these findings and calculations \footnote{Only the exact gradient derivation for the exact propagator is similar to the calculations in Ref.~\cite{de2011second}.} by first optimizing a minimal two-level Landau-Zener (LZ) problem and then a nine-level transmon system (Sec.~\ref{sec:numerical}).
In both instances, we attain the performance hierarchy qualitatively captured in Fig.~\ref{fig:landscape}.
We then show that this trend is exponentially monotonic in the face of more complex and larger systems. 
As the Hilbert space dimension scales exponentially in the number of constituents, this becomes especially relevant when, e.g., the system size enters the many-body regime where exact diagonalization, exact propagation, and associated exact derivatives are completely outside the realm of numerical feasibility. 
Finally, we show that the results generalize well to scenarios with more than one control (Sec.~\ref{sec:generalization})
and touch on a few pertinent discussion points (Sec.~\ref{sec:discussion}). 


\section{Exact Derivatives for \\ Quantum Optimal Control}
\label{sec:exactderivatives}

\subsection{Formulation of Unitary Control Tasks}
In quantum optimal control we seek to dynamically steer some quantum mechanical process in a controlled way such as to maximize a desired physical yield. For unitary evolution, any such task can be encoded as a minimization over an appropriate cost functional $J[\U(T;0)]$ where 
\begin{align}
\U(T;0) = \mathcal{T} \exp\bigg(-i \int_{0}^{T} \H(t) \mathrm{d}t\bigg), \label{eq:UOrdered}
\end{align}
is the time evolution operator in units where $\hbar = 1$ from time $t=0\rightarrow T$, $\mathcal T $ denotes time ordering, 
and $\op H$ is the system Hamiltonian carrying some generic time dependence.
The cost functional is a purely mathematical and malleable construct that quantifies our desired set of success criteria or goals. 
Quite often these not only include the desired quantum dynamics but also experimental constraints.
The individual criteria are typically represented by their own cost functional, $J_i$, and the total cost composed by $J=\sum_i J_i$. 
Typically one will define the cost such that $J \geq 0$ and $J=0$ is then guaranteed to be a global minimum.
We briefly return to different potential choices of $J_i$ at the end of this section.
Minimizing $J$ thus instructs us how to feasibly achieve the desired dynamics under the given experimental constraints.

The manipulatory access to the system dynamics in Eq.~\eqref{eq:UOrdered} is through a set of \textit{control} parameters in the Hamiltonian. To preserve clarity of the presentation we initially consider the case of a single generic control, $u(t)$, 
and we may without loss of generality separate the system Hamiltonian as 
\begin{align}
\H = \H(t,u(t)) = \Hd(t) + \Hc(t,u(t)). \label{eq:generalH}
\end{align}
In Sec.~\ref{sec:generalization} we extend our analyses to more than one control and show that the generalization remains feasible.
The drift Hamiltonian $\Hd(t)$ represents parts of the system dynamics that is uncontrollable.
The control Hamiltonian $\Hc(t,u(t))$, on the other hand, depends on the control $u(t)$ and grants us mandate to steer the dynamics. 
As example, for a single-particle system the drift could be the kinetic energy $\Hd = \op T$ and the control Hamiltonian the potential profile $\Hc=\op V(x,u)$.

Similarly to the cost $J$, it is gainful to think of $u(t)$ as a malleable mathematical object that is tied to some physical quantity $p(t)$ in the system such as the intensity, trap center, or frequency of a laser that ultimately affects the system.
In this general framing we can write $p(t) = g(u(t))$ where $g$ is some suitable differentiable function
and this allows a certain degree of modeling convenience.
For example, one might choose $g(u) = a\cdot u$ for numerical reasons where $c$ is an appropriate scaling constant \cite{nocedal2006numerical} such that appropriate values of $u$ is of order 1. 
Other examples include letting $g$ be a shifted and scaled sigmoid or arctan function \cite{sorensen2019qengine} such that $p$ is bounded within a range of values or letting $g$ be a so-called transfer function such that $p$ respects finite electronic response times \cite{sorensen2018quantum}. This provides e.g. alternative measures for incorporating experimental constraints without the use of additional cost functionals. 

Notice that our definition of the control Hamiltonian in Eq.~\eqref{eq:generalH} subsumes the control itself which is more general than the pervasive bilinear assumption, i.e. $ \Hc(t,u(t))\rightarrow u(t) \Hc(t)$. 
Linear physical dependences do indeed occur in many systems e.g. in the form of spin couplings which can be subject to control. 
Nevertheless, assuming control linearity precludes both the use of nonlinear $g$ as defined above as well as treatment of systems that are inherently not linear in the physical parameter. Examples of the latter include position-controlled Gaussian potential profiles $V(x,u)  \propto \exp(2(x-u)^2/\sigma^2)$ where $\sigma$ is a width \cite{weitenberg2011quantum} or phase-controlled optical lattices $V(x,u) \propto \cos^2(kx+u)$ where $k$ is a wave number \cite{mandel2003coherent}. 
For these reasons and since the derivations presented here do not depend on it we do not impose linearity assumptions
and simply note that this limiting case can always be taken at the end. \\

For numerical --- and as we shall see, analytical --- convenience it is natural to discretize time in regular $\dt$ intervals 
\begin{align}
t \in [t_1,t_2,\dots,t_{\nt}] &= [0, \dt, \dots, T], \quad  t_j = (j-1)\dt,
\end{align}  
with time indices denoted as subscripts. Physical quantities evaluated at these grid points are similarly denoted by $u_n = u(t_n)$ and $\H_n = \H(t_n,u_n)$ and similar for $\Hc_n$ and $\Hd_n$.
This leads to a product of time evolution operators
\begin{subequations}
\label{eq:Uproductapprox}
\begin{align}
&\U(T;0) \approx \prod_{j=1}^{\nt-1} \U_j = \U_{\nt-1}\dots \U_2\U_1, \label{eq:timeevolution} 
\end{align}  
where 
\begin{align}
\U_n = \exp\bigg(-i\int_{t_n}^{t_n+\delta t} \H(t,u(t))\,\mathrm{d}t\bigg), \label{eq:Uint}
\end{align}
\end{subequations}
is the propagator across the time interval $[t_n, t_{n+1}] = [t_n, t_n+\delta t]$.
As seen later, the error made with respect to Eq.~\eqref{eq:UOrdered} is given by $\nt-1$ first-order Suzuki-Trotter expansions and vanishes when $\dt\rightarrow 0 \leftrightarrow \nt \rightarrow \infty$ or if $[\H(t_j), \H(t_i)]=0$ for all $t_j$ and $t_i$. 
We return to the computation of $\U_n$ in Sec.~\ref{sec:landscapes} and for now simply note that $\U(T;0)$ depends on the discretized control vector $\vec u= (u_1,\dots,u_{\nt})^T$.


The optimal control task thus consists in finding appropriate control vector(s) $\vec u$ that correspond to local and hopefully global minima in the \textit{control landscape} defined by the cost
\begin{align}
J(\vec u)=J(\U_{\nt-1}\dots \U_2\U_1). \label{eq:J}
\end{align}
There are a plethora of techniques and prescriptions for maneuvering the landscape in search of such minima.

\subsection{Derivative-Based Local Optimization}
In this paper we focus on derivative-based local optimization methodologies characterized by making informed decisions in traversing the control landscape $\vec u^{(k)} \rightarrow \vec u^{(k+1)}$ using local information at iteration $k$ about the landscape topography. In prototypical linesearch-based updates of the form
\begin{subequations}
\begin{align}
&\vec u^{(k+1)} = \vec u^{(k)} + \alpha^{(k)} \vec p^{(k)}, \hspace{1cm}\alpha^{(k)} \in \mathbb{R}^+, \\
J(&\vec u^{(k+1)}) \leq J(\vec u^{(k)})
\label{eq:linesearch}
\end{align}
\end{subequations}
the search direction $\vec p^{(k)}$ is calculated from the current local gradient (e.g. steepest descent, conjugate gradient, quasi-Newton directions) and possibly also the Hessian (e.g. Newton direction). 
It is helpful to abstractly depict landscapes as in Fig.~\ref{fig:landscape} where the colormap denotes cost value $J(\vec u)$, each dot is an iterate $\vec u^{(k)}$ and the line connecting it to the next is the step $\alpha^{(k)} \vec p^{(k)}$. 
The step size $\alpha^{(k)}$ is in practice determined by an inexact linesearch \cite{nocedal2006numerical} and values largely depend on both the chosen linesearch algorithm, search direction, problem scaling, and how close to a minimum the iterate is. 
Far from the minimum, as is typical for the initial iterate, it can be on the order of hundreds or thousands after which it usually becomes of order one or smaller as it approaches convergence.

Although the functional form of the cost $J$ in Eq.~\eqref{eq:J} depends on the particular unitary task \cite{schirmer2011efficient} --- e.g. gate synthesis, state transfer, or maximization of a given observable and whether a pure state or density matrix description is considered, etc., --- they all lead to the same principal form for the control derivative calculations, specifically
\begin{subequations}
\label{eq:gradhess}
\begin{align}
\mathrm{Gradient} \sim&\hspace{0.7cm}\frac{\partial}{\partial u_n} \left( \U_{\nt-1}\dots
\U_2 \U_1\right), \label{eq:grad} \\
\mathrm{Hessian}\sim&\hspace{0.05cm}\frac{\partial^2}{\partial u_n \partial u_m}\left( \U_{\nt-1}\dots 
\U_2 \U_1\right). \label{eq:hess}
\end{align} 
\end{subequations}
Accurate control derivatives are paramount in successfully traversing the optimization landscape, since inaccuracies or willful approximations yield poor search directions and may significantly slow down, altogether prevent convergence, and/or limit the achievable fidelity c.f. exact landscape in Fig.~\ref{fig:landscape}. 
Out of the standard search directions, the steepest descent direction is well-known to have the comparatively worst general properties such as zig-zag iterate trajectories and a linear convergence rate \cite{nocedal2006numerical}. This is irrespective of the exactness of the gradient and inaccuracies will make this choice even more unattractive. 
The much more theoretically sound exact Newton direction exhibits quadratic convergence and a step size of $\alpha^{k}\rightarrow 1$ when approaching a minimum \cite{nocedal2006numerical}. It is, however, typically several orders of magnitude more expensive to construct than a steepest descent direction 
because it relies on both the gradient and the Hessian. Inaccuracies in either derivative thus negate any of the potential gains from the extra computational effort.
The class of quasi-Newton methods is generally accepted to be the most general-purpose class of search directions and exhibit super-linear convergence \cite{nocedal2006numerical}. 
For example, the \textsc{bfgs} direction is initialized as a steepest descent step and then builds up an approximation for the Hessian based on the gradients recorded in each iteration. 
If the individual gradients are inaccurate this error will accumulate in each iteration and make the Hessian approximation unreliable as the optimization progresses. 
This will in turn manifest as increasingly poor search directions that ultimately leads the 
optimization irreparably astray as depicted in Fig.~\ref{fig:landscape}. 
This is referred to as the ``slowdown'' problem in Ref.~\cite{de2011second}.

These considerations are generic and not limited to the field of quantum control. 
We calculate the exact derivatives that surpass these issues in Sec.~\ref{sec:analytical} and verify the calculations and claims above by numerical demonstration in Sec.~\ref{sec:numerical}. 
For specificity, we consider the case of pure state transfer, $\ket{\psio} \rightarrow \ket{\psitgt}$, ideally obtaining unit fidelity given by
\begin{align}
F&=|\braket{\psitgt| \psi(T)}|^2 = |\braket{\psitgt| \U(T;0) |\psio}|^2 \label{eq:F}, 
\end{align}
where $\ket{\psitgt}$ is the target state and $\ket{\psio}=\ket{\psi_1}$ is the initial state. The state $\ket{\psi(T)} = \ket{\psi_{\nt}}= \U(T;0) \ket{\psio}$ at final time $T$ is produced by stepwise evolution according to $\ket{\psi_{n+1}}=\U_n\ket{\psi_n}$. 
The associated fidelity cost is $J_F=(1-F)/2$.  
It will be convenient to write the fidelity as $F=\overlap^* \overlap$ and thus the 
fidelity cost and its derivatives as
\begin{subequations}
\label{eq:derivativeequations}
\begin{align}
J_F &=  \frac{1}{2}\big( 1- \overlap^* \overlap\big), \\
\frac{\partial J_F\ut{}}{\partial u_n} &= - \Re\left(\overlap^* \frac{\partial o}{\partial u_n}\right), \label{eq:gradientJF} \\
\frac{\partial^2 J_F\ut{}}{\partial u_n \partial u_m} &= -\Re \left( \left(\frac{\partial \overlap}{\partial  u_m}\right)^*  \frac{\partial \overlap}{\partial u_n} + \overlap^*  \frac{\partial^2 \overlap}{\partial u_n \partial u_m} \right), \label{eq:hessianJF} 
\end{align}
with $\overlap$ being the overlap/transfer amplitude with derivatives
\begin{align}
\overlap &=\Braket{\chi_{\nt}|\psi_{\nt}} = \Braket{\chi_{\nt}| \U_{\nt-1}\dots \U_n \dots\U_1   |\psi_1}, \label{eq:overlap}\\
\frac{\partial o}{\partial u_n} &= \Braket{\chi_{\nt}  |  \hspace*{0.65cm}\frac{\partial}{\partial u_n}  \left(\U_{\nt-1}\dots \U_n\dots \U_1\right)  | \psi_1}, \label{eq:doverlap}\\
\frac{\partial^2 \overlap}{\partial u_n \partial u_m} &= \Braket{\chi_{\nt}  |  \frac{\partial^2}{\partial u_n \partial u_m}  \left(\U_{\nt-1}\dots \U_n\dots \U_1\right)  | \psi_1} \label{eq:ddoverlap},
\end{align}
\end{subequations}
where we introduced an auxiliary state $\ket{\chi_{\nt}} = \ket{\psitgt}$ with $\ket{\chi_{n}} = \U_{n}^\dagger \ket{\chi_{n+1}}$.
The overlap derivatives is on the form of Eqs.~\eqref{eq:gradhess} as anticipated.  
The Hessian matrix is of course symmetric and allows reuse of gradient elements. 
To obtain numerically implementable expressions for the gradient and Hessian in Eqs.~\eqref{eq:gradientJF}--\eqref{eq:hessianJF} our task is now to analytically calculate Eqs.~\eqref{eq:doverlap}--\eqref{eq:ddoverlap}.
These calculations will clearly depend on the details of the propagator $\U_n$.

We stress that although the particular functional form for $J=J_F$ lead to ``specialized'' derivatives in Eqs.~\eqref{eq:derivativeequations}, dictated by the chain-rule, our evaluations of Eqs.~\eqref{eq:gradhess} are general which always constitute the by-far largest numerical effort. Thus, obtaining exact derivatives for any other unitary control task mentioned above --- e.g. gate synthesis, dynamics described by density matrices, etc. --- is a trivial extension by appropriately applying the chain-rule to the encoding functional. 
We therefore restrict our attention to the pure state transfer formulation in the remainder of this paper.
For completeness we also include derivatives for common control regularization terms in Appendix~\ref{app:derivatives}. 
These cost augmentations are as mentioned earlier often introduced for experimental reasons and they do not depend on the numerical propagation scheme.

\subsection{Suzuki-Trotter Expansions}
\label{sec:SuzukiTrotter}
\newcommand{\cst}{\alpha}
\newcommand{\cstt}{\frac{\alpha}{2}}
To set the stage for the following section we briefly recall a few ubiquitous Suzuki-Trotter expansions for the operator exponential. 

The exponential of the operator $\op X$ or its matrix representation is defined in terms of its Taylor series
\begin{align}
e^{\cst \op X}  = \sum_{k=0}^{\infty}\frac{\left(\cst \op X\right)^k}{k!} = 1+ \cst  \op X+ \frac{1}{2}\cst^2  \op X^2+ \mathcal O(\cst^3), \label{eq:expm}
\end{align}
where $\cst$ is a scalar and $\mathcal O(\cst^3)$ denotes terms of order $\cst^3$ or higher. 
Operator exponentials appear in many scientific contexts and the literature surrounding its explicit and efficient evaluation is quite extensive \cite{moler2003nineteen}. 
Let us assume that additional structure is present, $\op X = \op A + \op B$, in which case the expansion reads
\begin{align}
e^{\cst (\op A+\op B)}  
&= 1+ \cst  (\op A+\op B) \notag \\
&+ \frac{1}{2}\cst^2  (\op A^2+\op A\op B+\op B\op A+\op B^2)  + \mathcal O(\cst^3), \label{eq:trotterexp1}
\end{align}
after performing the square. 
Properties of $\op A$ and $\op B$ typically ensure that individual evaluation of $e^{\cst \op A}$ and $e^{\cst \op B}$ is much simpler than the composite $e^{\cst (\op A + \op B)}$.
For example, if $\op A$ is represented in a diagonal basis by the $N\times N$ matrix $\mat A$ then 
the matrix exponential is also diagonal with elements $(e^{\cst \mat A})_{n,n} = e^{\cst \mat A_{n,n}}$ for all $n=1,\dots,N$.

Motivated by this fact and the rules for scalar exponentials, evaluating the product $e^{\cst \op A} e^{\cst \op B}$ using Eq.~\eqref{eq:expm} twice leads to the simplest, first-order Suzuki-Trotter expansion \cite{hatano2005finding} by comparing to Eq.~\eqref{eq:trotterexp1}
\begin{subequations}
	\label{eq:firstordertrotter}
\begin{align}
e^{\cst \op A}e^{\cst \op B}  &= 
1+\cst(\op A + \op B) + \frac{\cst^2}{2}(\op A ^2 + 2\op A\op B + \op B^2) + \mathcal O(\cst^3) \notag \\
&=e^{\cst (\op A+\op B)}  + \mathcal O(\cst^2).
\end{align}
The first-order expansion evidently has an error scaling 
\begin{align}
e^{\cst (\op A+\op B)}  - e^{\cst \op A}e^{\cst \op B}  = \frac{\cst^2}{2}[\op B,\op A] +  \mathcal O(\cst^3) = \mathcal O(\cst^2), 
\end{align}
\end{subequations}
which depends on the commutator $[\op B,\op A]$.
In fact it is well known that $e^{\cst (\op A+\op B)}=e^{\cst \op A}e^{\cst \op B}$ is {exact} if $[\op B,\op A] = 0$. 
Suppose we instead considered an ansatz on the form $e^{\cst \op A_1}e^{\cst \op B}e^{\cst \op A_2}$ where 
\begin{subequations}
	\label{eq:secondordertrotter}
\begin{align}
\op A = \op A_1+\op A_2 +\mathcal{O}(\cst^2). \label{eq:conditationA1A2}
\end{align}
By applying Eq.~\eqref{eq:expm} three times we obtain 
\begin{align}
&e^{\cst \op A_1}e^{\cst \op B}e^{\cst \op A_2} = 1 + \cst(\op A_1 + \op A_2 + \op B) \notag \\
&+ \frac{\cst^2}{2}(\op A_1^2 + \op A_2^2 + 2(\op A_1\op A_2 +\op A_1 \op B+ \op B \op A_2) + \op B^2) + \mathcal{O}(\cst^3).
\end{align}
This Suzuki-Trotter expansion has error
\begin{align}
e^{\cst (\op A +\op  B)} - e^{\cst \op A_1}e^{\cst \op B}e^{\cst \op A_2} &= \frac{1}{2}\cst^2([\op A_2,\op A_1] + [\op B,\op A_1-\op A_2]) \notag \\
&+\mathcal O(\cst^3),
\end{align}
\end{subequations}
which is also to first-order $\mathcal O(\cst^2)$ for arbitrary $\op A_1$ and $\op A_2$, but to second-order $\mathcal O(\cst^3)$ when 
\begin{subequations}
	\label{eq:Aconditions}
	\begin{align}
	[\op A_1,\op A_2]&=0. \label{eq:Acondition12comm}\\
	\op A_1-\op A_2 &= \mathcal O(\cst), \label{eq:Acondition12}
\end{align}
\end{subequations}
More generally it is possible to systematically construct Suzuki-Trotter expansion variants of arbitrarily high order 
by considering e.g. the ansatz $e^{p_1\cst \op A}e^{p_2\cst \op B}e^{p_3\cst \op A}e^{p_4\cst \op B} \dots e^{p_{M}\cst \op B} + \mathcal{O}(\cst^{m+1})$ and choosing suitable coefficients $p_i$ \cite{hatano2005finding}.
Although arbitrarily low error is an enticing prospect it also entails more computational time. 
Since in our context $\alpha = -i \dt$ represents a small time step it is sufficient to henceforth only consider expansions such as the ones above. 


\subsection{Dynamics and Optimization Landscapes}
\label{sec:landscapes}
We now return to the numerical evaluation of Eqs.~\eqref{eq:Uproductapprox}.  
To compute the propagator in Eq.~\eqref{eq:Uint} we must first perform the integral in the exponential. 
We consider numerical quadratures based on the left-point rectangle rule and the trapezoidal rule,
\begin{align}
\label{eq:integration}
\int\displaylimits_{t_n}^{t_n+\delta t} \H(t,u(t))\, \mathrm{d}t \approx
\dt
\begin{cases}
\hat{\oline{H}}_n  &\text{(trapezoidal)}, \\
\H_n  &\text{(rectangle)},\\
\end{cases}
\end{align}
where $\H_n = \H(t_n, u_n)$ and $\hat{\oline{H}}_n = \frac{1}{2}(\H_{n+1} + \H_{n})$.
If the underlying time dependence is assumed continuous the trapezoid and rectangle approximations have integration errors $\mathcal O(\dt^3)$ and $\mathcal O(\dt^2)$, respectively. 
If the time dependence is assumed piecewise constant the rectangle approximation is exact. 
Both rules thus satisfy the condition in Eq.~\eqref{eq:conditationA1A2}.

Depending on the choice of quadrature we will refer to the \textit{exact} exponentiation propagators as 
\begin{subequations}
	\label{eq:propagators}
	\begin{align}
	\U\ut{Ex_1}_n &=  \exp\bigg({-i \hat{\oline{H}}_n\dt}\bigg), \label{eq:exactPropTrap} \\
	\U\ut{Ex_2}_n &=  \exp\bigg({-i\H_{n}\dt}\bigg), \label{eq:exactProp} 
	\end{align}
	and their corresponding \textit{Suzuki-Trotter} (or \textit{Trotterized}) propagators as, respectively, 
	\begin{align}
	\U_n\ut{ST_1} &= \U_{n+1}^{c/2} \U_n^d \U_{n}^{c/2} \approx  \U\ut{Ex_1}_n,   
	\label{eq:STProp}  \\
	\U_n\ut{ST_2} &= \U_{n}^{c/2}\hspace{0.1cm} \U_n^d \U_{n}^{c/2} \approx  \U\ut{Ex_2}_n, 
	 \label{eq:STProp2} 	
	\end{align}
	with definitions for the control and drift exponentials
	\begin{align}	
	\U_n^{c/2} &\equiv \exp\bigg({-i\H_{n}^c \dt/2}\bigg),  \\
	\U_n^d &\equiv 
	\begin{cases}
	\exp\bigg({-i\hat{\oline{H}}_n~\!\!\!\!\!^d \dt}\bigg), \quad &\text{for } \mathrm{ST_1}, \\
	\exp\bigg({-i\Hd_n \dt}\bigg), \quad &\text{for } \mathrm{ST_2}.
	\end{cases}
	\label{eq:driftexponential}
	\end{align}
\end{subequations}
The operator splitting $\U_n\ut{ST_2}$ is achieved by utilizing Eqs.~\eqref{eq:secondordertrotter} with $B=\Hd_n$ and $\op A_1=\op A_2 = \Hc_n/2$ and always has Trotterization error $\mathcal{O}(\dt^p)$ with $p=3$ according to Eqs.~\eqref{eq:Aconditions}. 
The operator splitting $\U_n\ut{ST_1}$ is achieved by letting $B=\hat{\oline{H}}_n~\!\!\!\!\!^d$, $\op A_1 = \Hc_{n+1}/2$, and $\op A_2 = \Hc_{n}/2$ and has Trotterization error $\mathcal{O}(\dt^{p})$ where $p=2,3$ depending on Eqs.~\eqref{eq:Aconditions}. 
For unrelated reasons we will later assume that the control Hamiltonians are diagonal which leaves only condition Eq.~\eqref{eq:Acondition12}. However, the precise Trotter error $p$ is not of crucial importance since $\dt$ must under all circumstances be small enough that the errors made in going from Eq.~\eqref{eq:UOrdered} to Eqs.~\eqref{eq:Uproductapprox} 
\footnote{This corresponds to $\nt-1$ applications of the first-order expansion in Eqs.~\eqref{eq:firstordertrotter}.}
and in integrating Eq.~\eqref{eq:integration} are small. 
What is much more important is that $\U_n\ut{ST_2}$ is fully local in $n$ whereas $\U_n\ut{ST_1}$ depends on both $n$ and $n+1$.
The resulting derivative calculations with respect to $u_n$ and final expressions are thus different. This underscores that the precise specification of the implementation is central for use in optimal control contexts. The derivatives should ``match'' the dynamics.

The local $\mathcal{O}(\dt^{p})$ Trotterization errors accumulate throughout the $N_t-1$ evolutions in Eq.~\eqref{eq:timeevolution} yielding an overall error $(\nt-1) \dt^p \approx (T/\dt)\dt^p = T \dt^{p-1} \sim \mathcal O(\dt^{p-1})$.
It is convenient to interpret this as an approximation error with respect to the exact landscape c.f. Fig.~\ref{fig:landscape},
\begin{align}
J_F\ut{ST}(\vec u) = J_F\ut{Ex}(\vec u) + \mathcal{O}_{J_F\ut{ST}}(\dt^{p-1}), \label{eq:landscape}
\end{align}
for $\textsc{e}\mathrm{x}=\textsc{e}\mathrm{x}_1, \textsc{e}\mathrm{x}_2$ and associated $\textsc{st}=\textsc{st}_1, \textsc{st}_2$. 
The granularity of $\dt$ determines how faithful the representation is 
and it follows that 
geometric entities for the \textit{same} point $\vec u$ are generally different in each landscape. This includes the height/cost value, derivatives, and thus also the search directions for optimization.
Importantly, the optimal controls associated with optima in the Trotterized landscapes at large finite $\dt$ may not correspond to optima in the exact landscapes, which is equivalent to the target not being obtained when propagating said controls using Eqs.~\eqref{eq:exactPropTrap}--\eqref{eq:exactProp}. 
As $\dt\rightarrow 0$, however, the Trotterized landscapes continuously deform into the exact landscapes, and below some sufficiently small finite $\dt$ they represent it with only negligible perturbations. 
The landscapes transitively inherit the numerical implementation properties of their associated propagator. \\

Numerically, the exact propagator corresponds to direct exponentiation of the Hamiltonian matrix, an operation that scales extremely poorly with increasing Hilbert space dimension $D_\mathcal{H}$.
On the other hand, the Trotterized propagators lend themselves more readily to a variety of very efficient, problem-dependent implementations through, e.g., the use of sparsity structures, and these have a much more benign Hilbert space scaling, extending its applicability far beyond the exact propagator approach.
For example, we shall assume that the control Hamiltonian is diagonal and this significantly boosts runtime performance.
This is because $\U^{c/2}_n$ is relatively cheap to represent and calculate since diagonal matrix exponentiation is the element-wise exponentiation of the diagonal. Additionally, the relatively very expensive calculations of $\U^{d}_n$ for all $n$ can be performed once and stored in memory or on the disk as they are by definition independent of the choice of $\vec u$. 
We can also write vectorized forms useful for state-transfer and unitary synthesis, respectively, as%
\begin{subequations}
\begin{align}
&\mat{\mathcal{U}}\ut{ST_1}_n \times \bm{\psi_n} = \vec{\mathcal{U}}^{c/2}_{n+1}\odot\left(\mat{\mathcal{U}}^d_n\times \left(\vec{\mathcal{U}}^{c/2}_{n} \odot\vec{\psi_n}\right)\right), \label{eq:vectorized}
\\&\mat{\mathcal{U}}\ut{ST_1}_n  = \mat{\mathcal{U}}^d_n \odot \left(\vec{\mathcal{U}}^{c/2}_{n+1} \times (\vec{\mathcal{U}}^{c/2}_{n})\ut{T} \right),
\end{align}%
\end{subequations}
where $\mat{\mathcal{U}}\ut{ST_1}_n$, $\mat{\mathcal{U}}^d_n$ are dense matrices, $\vec{\mathcal{U}}^{c/2}_{n}$, $\vec{\psi_n}$ are a vectors,
$\times$ ($\odot$) denotes regular (element-wise) matrix multiplication, and $\ut{T}$ is transposition.
These numerical techniques are the original reasons for approximating the dynamics. 
In the next section we show that also more subtle and important simplifications occur when calculating the exact derivatives.
\\

\section{Analytical Results}
\label{sec:analytical}
Detailed calculations of the analytically exact derivatives stated in Eqs.~\eqref{eq:derivativeequations} using the propagators in Eqs.~\eqref{eq:propagators} are given in Appendix~\ref{app:derivatives}. 
Here we focus on the central equations and contexts. 
Of the exact propagators, we consider for simplicity only $\U_n\ut{Ex} = \U_n\ut{Ex_2}$ given by Eq.~\eqref{eq:exactProp} 
from hereon. \\

By defining the recursive commutator 
\begin{align}
[X,Y]_k &= [X,[ X,Y]_{k-1}] , \qquad [X,Y]_0 = Y, \label{eq:recursivecomm}
\end{align} 
a central calculation shows that 
\begin{align}
\frac{\partial \U\ut{Ex}_n}{\partial u_n} &= \U\ut{Ex }_n   (-i\dt)  \sum_{k=0}^{\infty}  \frac{i^k \dt^{k} }{(k+1)!} [\H_n,\frac{\partial \Hc_n}{\partial u_n}]_k. \label{eq:dUdu}
\end{align}
where $\partial \H_n^{c} /\partial u_n = \partial\H_n/\partial u_n$ is the control derivative Hamiltonian. 
The presence of the infinite sum means that in the context of optimization it may not be desirable to use exact time evolution even if it is readily available.
In numerical application, the summation continues until machine precision or to a desired accuracy, corresponding to some $k\st{max}$. This is necessary because the Hamiltonian and its control derivative generally do not commute, $[\H_n, \frac{\partial \H_n^{c}}{\partial u_n}] \neq 0$, and the recursive commutator is not guaranteed to terminate.
A few examples include $\op H = \op \sigma_x  + u(t)\cdot \op \sigma_z$ where $\op \sigma_i$ are Pauli spin-$\frac{1}{2}$ operators, $\op H = \op T + \op V(u(t))$ where $\op V$ ($\op T$) is the potential (kinetic) energy operator for a single particle, or the Bose-Hubbard Hamiltonian $\op H = -\op J + \op U(u(t))$ where $\op J$ ($\op U$) is the tunneling (on-site interaction) operator. 

The propagator derivatives for the Trotterizations in Eq.~\eqref{eq:STProp}--\eqref{eq:STProp2} contain an infinite series of the same structure as in Eq.~\eqref{eq:dUdu}, but with $\H_n \rightarrow \H_n^{c}$ in the first argument of the recursive commutator. 
Thus, by additionally assuming that the control Hamiltonian is expressed in its diagonal representation 
we find
\begin{subequations}
\begin{align}
&\frac{\partial}{\partial u_n}\left(\U_{n}\ut{ST_1} \U_{n-1}\ut{ST_1}\right)  =(-i\dt) \cdot \U_{n}\ut{ST_1}  \frac{\partial \H_n^{c}}{\partial u_n}\U_{n-1}\ut{ST_1}, \\
&\frac{\partial \U_{n}\ut{ST_2}}{\partial u_n}  = -\frac{i\dt}{2} \left( \frac{\partial \H_n^{c}}{\partial u_n} \U_{n}\ut{ST_2}  + \U_{n}\ut{ST_2}  \frac{\partial \H_n^{c}}{\partial u_n}\right). \label{eq:UST2deriv}
\end{align}
\end{subequations}
since the series terminate exactly after $k\st{max} = 0$.
This huge simplification occurs because two diagonal matrices always commute, $[\H_n^c,\frac{\partial \H_n^{c}}{\partial u_n}]_k = \frac{\partial \H_n^{c}}{\partial u_n} \cdot \delta_{0,k}$.
Incidentally, in many cases the ``natural'' basis states for computations are already the ones that diagonalize $\H_n^{c}$, e.g. 
spin eigenstates ($\ket{\uparrow}$,$\ket{\downarrow}$), position eigenstates ($\ket{\vec x}$), or site-occupation eigenstates ($\ket{n_i}$), respectively, for the Hamiltonians mentioned above. 

Inserting the results in Eqs.~\eqref{eq:dUdu}--\eqref{eq:UST2deriv} into Eqs.~\eqref{eq:derivativeequations} we obtain the gradient elements
\begin{subequations}
	\label{eq:gradients}
\begin{align}
\frac{\partial J_\F \ut{Ex}}{\partial u_n} 
&= \Re \left(i \overlap^*  \Braket{\chi_n |  \frac{\partial\H_n^{c}}{\partial u_n} | \psi_{n}  }\right)  \dt  + \mathcal O_{\nabla J_F\ut{Ex}}(\dt^2),  \label{eq:gradJexact} \\
\frac{\partial J_\F\ut{ST_1}}{\partial u_n} 
&= \Re \left(i \overlap^*\Braket{\chi_{n} | \frac{\partial\H_n^{c}}{\partial u_n}| \psi_{n}  } \right)\dt, \label{eq:exactGradientST} 
\\
\frac{\partial J_\F\ut{ST_2}}{\partial u_n}  &= \Re \bigg(\frac{i \overlap^*}{2} \bigg\{\Braket{\chi_{n+1} | \frac{\partial\H_n^{c}}{\partial u_n} | \psi_{n+1}} \notag\\
&\hspace{1.95cm}+ \Braket{\chi_{n} | \frac{\partial\H_n^{c}}{\partial u_n} | \psi_{n}}\bigg\} \bigg)\dt, 
\label{eq:exactGradientST2} 
\end{align}
\end{subequations}
The states in each equation are understood to be evolved according to the propagation scheme denoted on the left hand side but this notational completeness is omitted here for brevity.
It is important to remember that these Trotter gradients assume that the control Hamiltonian $\Hc$ is diagonal.
The remainder term for the exact propagator inherits the infinite series of Eq.~\eqref{eq:dUdu}, 
\begin{align}
\mathcal O_{\nabla J_F\ut{Ex}}(\dt^2) \propto \Braket{\chi_{n} | \left(  \sum_{k=1}^{\infty}  \frac{i^k \dt^{k} }{(k+1)!} [\H_n,\frac{\partial\H_n^{c}}{\partial u_n}]_k \right)| \psi_{n}}\dt,
\label{eq:exactderiv_tail}
\end{align}
where the first-order approximation in Eq.~\eqref{eq:gradJexact} is the $k=0$ term.

Thus, the gradients for $J\ut{Ex}$ and $J\ut{ST_1}$ are on the same form \footnote{Except at the endpoints, see Appendix~\ref{app:derivatives}} only to first order in $\dt$:  
whereas $\nabla J_\F\ut{ST_1}$ is \textit{analytically exact} with just the $\dt$ term, $\nabla J_\F\ut{Ex}$ entails an expensive remainder term beyond the first-order approximation. 
This is noteworthy and nontrivial since, while $\ut{ST_2}$ is the most commonly encountered type of Trotterization, $\nabla J_\F\ut{Ex}$ and $\nabla J_\F\ut{ST_2}$ do \textit{not} coincide even to first order. Generally, neither do the derivatives for other dynamical approximations such as Krylov-Lanczos, finite Taylor expansions, Crank-Nicolson, and Chebyshev schemes \cite{beerwerth2015krylov,dalgaard2020hessian,fehske2009numerical}.
Nevertheless, the gradient $\nabla J_\F\ut{ST_2}$ is only slightly more involved than $\nabla J_\F\ut{ST_1}$ due to two sum terms and this is negligible compared to the computational effort of the time evolution and overlap calculations can be reused. 
Preference towards either may therefore rely on which of the integration quadratures in Eq.~\eqref{eq:integration} is most appropriate in a given situation. 
Similar calculations and arguments apply to the analytically exact Hessians, but in this case, the expressions for $\nabla^2 J_\F\ut{Ex}$ and $\nabla^2 J_\F\ut{ST_2}$ are much more complicated than $\nabla^2 J_\F\ut{ST_1}$ as shown in Appendix~\ref{app:derivatives}. \\

An alternative way of calculating the exact propagator derivatives is through the auxiliary matrix method \cite{goodwin2015auxiliary,goodwin2016modified}. Focusing on the first derivative, the relation
\begin{align}
\begin{pmatrix}
\U_n\ut & \frac{\partial \U_n}{\partial  u_n}  \\ 
0 & \U_n 
\end{pmatrix}
=
\exp
\left[
-i
\begin{pmatrix}
\H_n & \frac{\partial\H_n{}}{\partial u_n} \\
0 & \H_n 
\end{pmatrix}
\dt
\right]
\label{eq:auxiliary}
\end{align}
allows extraction of both $\U_n$ and $\frac{\partial \U_n}{\partial  u_n}$ by explicitly calculating the right hand side $2 \times 2$ block matrix exponential. The expression can be augmented to a $3 \times 3$ block matrix to also include the Hessian. This approach provides exact derivatives while elegantly circumventing the cumbersome commutator series in Eq.~\eqref{eq:exactderiv_tail} associated with the exact propagator, but the required $2 D_\mathcal{H}$ (or $3 D_\mathcal{H}$ with Hessian) square matrix exponentials becomes similarly expensive. 

What happens if we combine the auxiliary matrix with our Trotterization with diagonal controls? 
The derivative of $u_n$-dependent propagators is e.g.
$\frac{\partial}{\partial u_n}\left(\U_{n}\ut{ST_1} \U_{n-1}\ut{ST_1}\right) =  \left(\U_{n+1}^{c/2} \U_{n}^d\right) \frac{\partial \U_{n}^{c}}{\partial u_n}  \left(\U_{n-1}^d \U_{n-1}^{c/2} \right)$
and we can then employ Eq.~\eqref{eq:auxiliary} with $ \U_n \rightarrow \U_n^{c}$ and $\H_n \rightarrow \H_n^c$. 
Since $\H_n^c$ is assumed diagonal, the right hand side exponent is a very sparse, \textit{almost} diagonal matrix with a single dense off-diagonal. Although very efficient numerical sparse solvers exist, the computation remains nontrivial, and it is always cheaper to instead utilize either Eqs.~\eqref{eq:exactGradientST}--\eqref{eq:exactGradientST2} in this case since the exponential itself is analytically trivialized and $\frac{\partial\H_n^{c}}{{\partial u_n}}$ is also on analytically closed form. 
\medskip

In Sec.~\ref{sec:discussion} we also discuss the results in comparison to the Krylov-Lanczos propagation scheme, another common type of approximate time evolution for quantum states.
Other types of techniques for extending simulation capabilities such as the reduced density matrix in e.g. \textsc{nmr} systems \cite{kuprov2007polynomially} will not be discussed further here. \\

\section{Numerical Results}
\label{sec:numerical}
To numerically verify the analytical conclusions and assertions made in Sec.~\ref{sec:analytical}, we initially examine the methods' performance capacities on two concrete state-transfer problems with dimensionality of $D_\mathcal{H} = 2$ and $D_\mathcal{H} = 9$, respectively. 
We then investigate the gradient evaluation time as $D_\mathcal{H}$ continually increases. 
In general, the particular parameters chosen for these studies are not central to the overall methodological conclusions and will only be discussed to the extent that they are relevant. 
The process duration $T$ for the presented results is chosen such that $T\st{min}^{F=1} \lesssim T \ll T\st{adiabatic}$ where $T\st{min}^{F=1}$ is the minimal duration that $F=1$ solutions exist and $T\st{adiabatic}$ is the adiabatic limit. 
We also briefly discuss the behavior at $0.5$, $0.75$, $1.25$ and $1.5$ times this $T$. 
All results were generated on a 2017 Macbook Pro laptop with 16\si{GB} RAM using a single 2.8\si{GHz} Intel Core i7 processor.

\subsubsection{Two-Level System}
\label{sec:twolevel}
We first consider the canonical LZ model
 with $\H_n = \H^d + \H_{n}^{c} = \frac{1}{2}\left( \op \sigma_x + u_n \cdot \op \sigma_z \right)$ and the state transfer $\ket{\psio} = \ket{\uparrow} \rightarrow \ket{\psitgt} = \ket{\downarrow}$. Employing the ``natural'' basis $\{\ket{\uparrow},\ket{\downarrow}\}$, this problem is already represented in the necessary control-diagonal form.
The reason for choosing this problem is twofold. 
First, it represents the smallest possible nontrivial type of problem ($D_\mathcal{H} = 2$).
Second, it has well-understood solutions \cite{hegerfeldt2013driving}
with a single, analytical $\pi$-pulse solution $u_n = 0$ at the minimal $F=1$ duration, $T^{F=1}\st{min}=\pi$, and remains solvable beyond this duration.
Yet despite of its simplicity the model remains prototypical even in the context of the many-body arena.
For example, many-body dynamics can in certain scenarios be thought of as a cascade of independent LZ transitions and 
similar characteristics between LZ- and some many-body control problems have been identified \cite{santoro2002theory,caneva2011speeding, caneva2009optimal}. \\

\begin{figure}
	\includegraphics[]{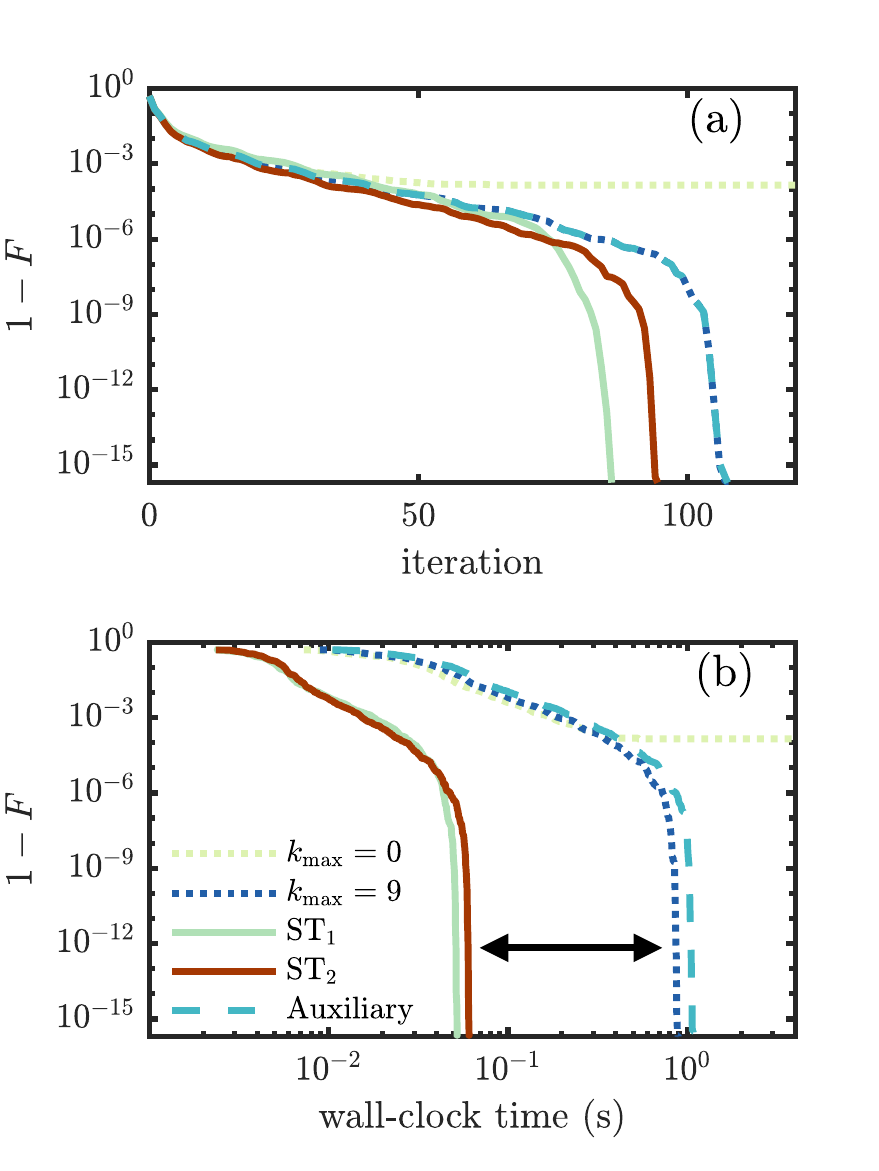}
	\includegraphics[]{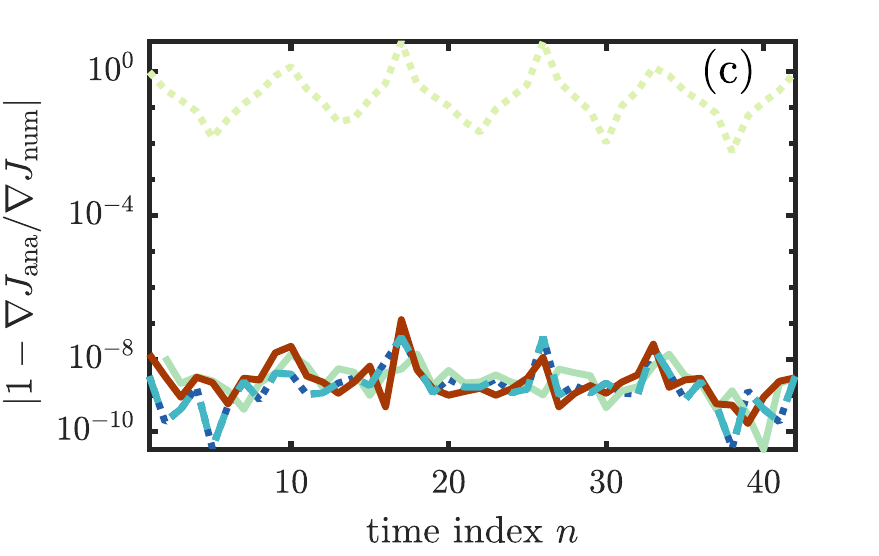}
	\caption{
		Demonstration of optimizing in exact- and Trotterized landscapes for the simple two-level LZ problem.
		The $1-F$ median as a function of (a): iteration and (b): wall-clock time. 
		The performance gap between the exact methods is roughly denoted by the horizontal arrow.
		(c): Analytical gradients $(\nabla J\st{ana})$ relative to their numerical central finite difference approximation $(\nabla J\st{num})$ with perturbation $\epsilon\st{pert} = \epsilon\st{mach}^{1/3}$ for a constant example control with $u_n = 5$. Note that the Auxiliary and $k\st{max} = 9$ gradients are identical to machine precision and thus yield (nearly) the same iteration trajectories, and, incidentally, nearly the same wall-clock time trajectories.
	}
	\label{fig:2level_example}
\end{figure}

We optimize the same 100 uniformly randomly generated seeds, $u_n = \mathrm{uniform}(-10,10)$, at $T=1.01\pi \gtrsim T^{F=1}\st{min}$ with $\dt = 0.075$ using the \textsc{bfgs} search direction implemented in \textsc{matlab}'s \texttt{fminunc} in five different scenarios: exact propagator with first-order ($k\st{max} = 0$) and exact ($k\st{max} = 9$ and auxiliary method) gradients, and both Trotterized propagators with exact gradients. 
Figure~\ref{fig:2level_example} shows the optimization results. Only 14 seeds did not converge to machine precision within 400 iterations when using exact gradients.

From the $1-F$ iteration median trajectories we find that unit fidelities to machine precision are easily obtainable only when utilizing any of the exact gradients whereas the inexact, approximate variant leads to poor results.
Solutions of the former exhibit very rapid convergence when they approach the optimum with a variance in the low tens for the number of iterations needed. 
Looking, however, at the optimization wall-clock time trajectories provides a definite performance hierarchy with negligible variance.
The mint and red trajectories are separated by more than an order of magnitude in computation time from the dark and light blue trajectories.
These medians are associated with the Trotterized and exact landscapes qualitatively shown in Fig.~\ref{fig:landscape}, respectively, and the separation is due to both the difference in propagation computation time and either the $k\st{max}>0$ tail in Eq.~\eqref{eq:exactderiv_tail} or the $2D_\mathcal{H} \times 2D_\mathcal{H}$ matrix exponential in Eq.~\eqref{eq:auxiliary}. 
As mentioned, the first-order gradient approximation $k\st{max}=0$ in the exact landscape performs significantly worse while also being slower compared to iterations in the Trotterized landscapes.  
Note however that its performance is decent up until around 50 iterations at which point it prevents convergence to unit fidelities by more than 10 orders of magnitude. 
This is because the Hessian approximation eventually becomes completely unreliable as discussed in Sec.~\ref{sec:exactderivatives}.

As a further verification of Eqs.~\eqref{eq:gradients}  we compare the analytical gradients to their respective central finite difference gradients
and find relatively close agreement. The latter are themselves associated with errors of order $\mathcal{O}_{\nabla J_\F}(\dt^2)$, and the relative differences are on the order $\sqrt{\epsilon\st{mach}}$ where $\epsilon\st{mach}=2.22 \cdot 10^{-16}$ is the machine precision for the double-precision floating-point format. 
We also quantitatively find that the first-order approximation is unsurprisingly poor.


The truncation parameter necessary for exact gradients is roughly bounded $\dt^{k\st{max}+1}/(k\st{max} + 1)! \lesssim \epsilon\st{mach}$. With decreasing $\dt$, the necessary $k\st{max}$ for exact gradients also decreases and the $k = 0$ term becomes increasingly dominant.  
Indeed, running the same optimizations as shown in Fig.~\ref{fig:2level_example} for $\dt = 0.025$, the $k\st{max} = 0$ optimization yields
final results that are 2--3 orders of magnitude better relative to $k\st{max}=0$ in Fig.~\ref{fig:2level_example} with $\dt=0.075$. At $\dt=0.01$ this first-order approximation is sufficient for finding machine precision unit fidelities.
That is, reducing the number of $k$ terms required for accurate gradients is traded off for increased computation time per iteration due to additional time evolutions.
Consequently, even though such $k\st{max}=0$ trajectories may now be sufficient in terms of final results, they are stretched to much higher wall-clock times than any of the results shown in Fig.~\ref{fig:2level_example}. \\

We also performed optimizations at durations $0.5$, $0.75$, $1.25$ and $1.5$ times $T=1.01\pi$. 
For both the smaller and larger durations we find empirically that the control landscape becomes very easy in the sense that only a few iterations ($<10$) are required for convergence and the optimized fidelities are nearly identical. 
This is not surprising since the problem is relatively simple even when the control is heavily constrained \cite{larocca2018quantum}.
Due to this simplicity and the associated low number of required iterations, the approximate gradient actually becomes competitive 
to the exact ones in the exact landscape in terms of wall-clock time at these durations. 
That is, convergence is achieved before the Hessian approximation becomes very unreliable.
This simplicity should not be expected for the majority of control problems, and a detailed study of this behavior and at intermediate $T$ is outside the scope of this paper. 
The performance gap to the exact gradient methods in the approximate landscape seen in Fig.~\ref{fig:2level_example} persists across all the different tested $T$ values regardless. 
Together with the discussion above this shows that the optimal choice of $k\st{max}$ is nontrivial and dependent on the other problem parameters. This suggests a methodological simplicity of the Trotter approach where gradient exactness is always ensured by just the first-order term.

\begin{figure}
	\includegraphics[]{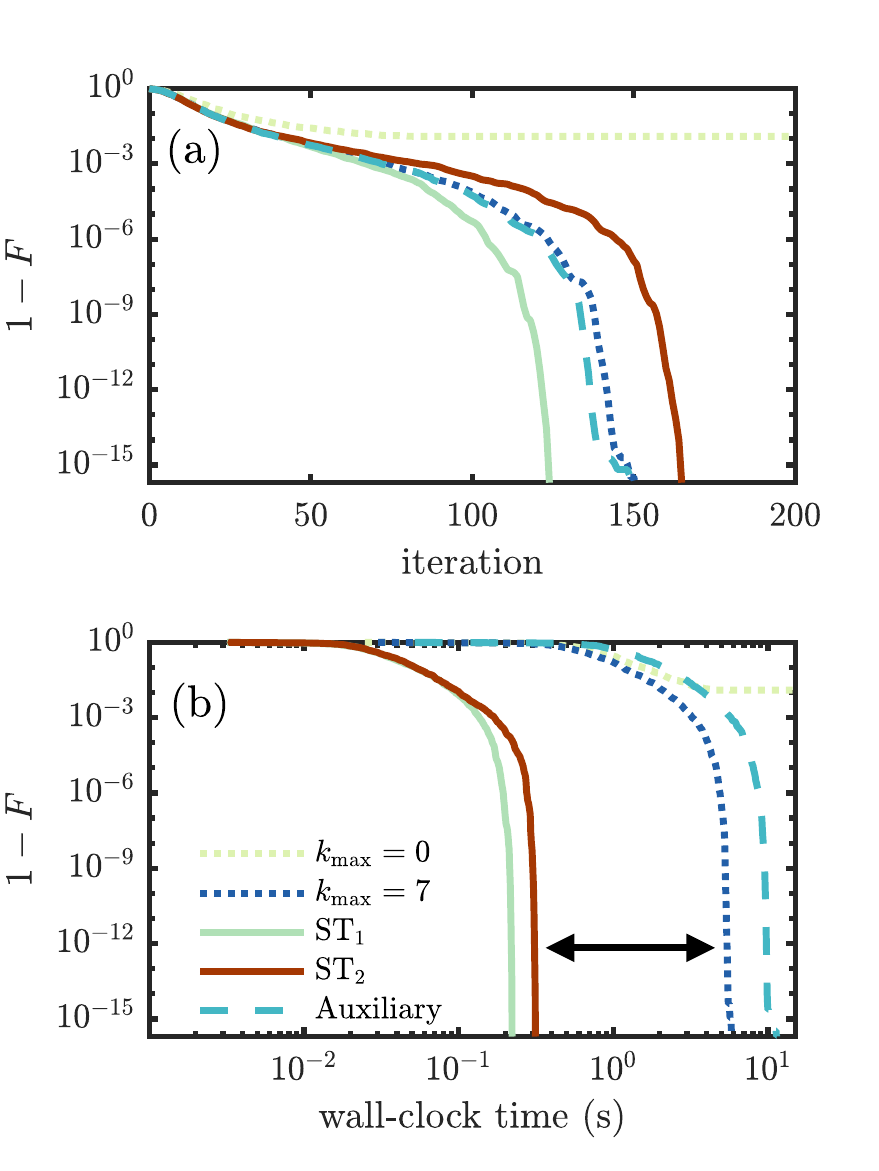}
	\includegraphics[]{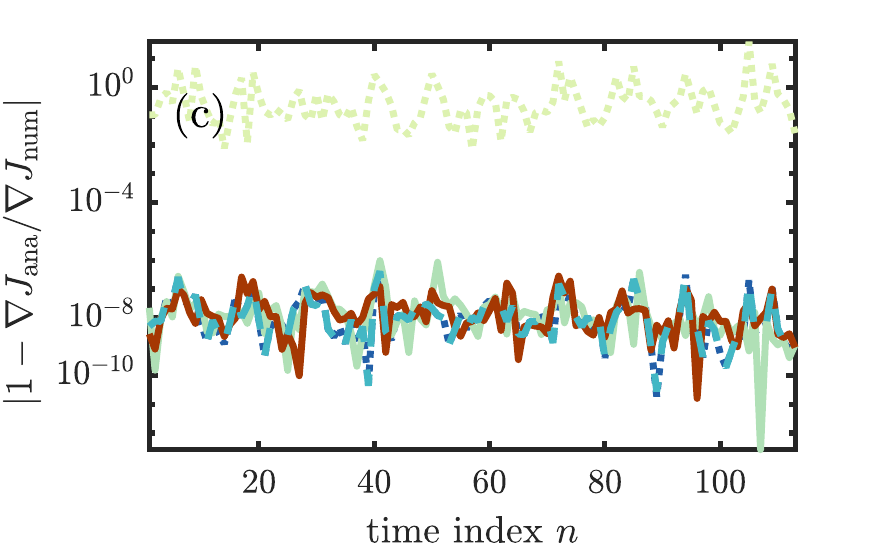}
	\caption{
		Same as Fig.~\ref{fig:2level_example} for 50 seeds, but for the transmon system Eqs.~\eqref{eq:transmonH}--\eqref{eq:basischange2} with $T=2.83\; (50\,\si{ns})$ and $\dt = 0.025\; (0.442\,\si{ns})$ given in nondimensionalized numerical units and \textsc{si}-units, respectively (energy is measured in units of $|\kappa| \approx 5.97 \cdot 10^{-27}\,\si{J}$ and time in $\hbar/|\kappa| \approx 17.7 \si{ns}$).
	}
	\label{fig:transmon_example}
\end{figure}

\subsubsection{Transmon System}
\label{sec:transmon}
We now turn to a second example of higher dimensionality ($D_\mathcal{H} = 9$), a superconducting transmon system with two-qutrit 
computational basis $\{\ket{0 0}, \ket{1 0}, \dots \ket{1 2},\ket{2 2} \}$ described by the Hamiltonian
\begin{align}
\H_n &= \left[\Delta \op b_1^\dagger \op b_1 +  \frac{1}{2}\sum_{j=1,2} \delta_j \op b_j^\dagger \op b_j (\op b_j^\dagger \op b_j - 1) + \kappa (\op b_1^\dagger \op b_2 + \op b_1 \op b_2^\dagger) \right] \notag\\
& + u_n(\op b_1^\dagger + \op b_1) = \Hd + \Hc_n, \label{eq:transmonH}
\end{align}
with the same parameter values as in Ref.~\cite{dalgaard2020hessian}, here with the relabeling $J\rightarrow \kappa$ to avoid ambiguity with the cost functional. We consider the state transfer $\ket{\psio} = \ket{10} \rightarrow \ket{\psitgt} = \ket{11}$, i.e. a single state mapping of a \textsc{cnot} gate in the qubit subspace $\{\ket{00}, \ket{01},\ket{10},\ket{11}\}$, without control constraints 
\footnote{
This process is expected to have a lower quantum speed limit than what we found for the full gate \cite{dalgaard2020hessian}. The full \textsc{cnot} could have been considered in the state transfer formulation by optimizing a composite cost, e.g. $J_F^{\ket{00} \rightarrow\ket{00}}+J_F^{\ket{01} \rightarrow\ket{01}} +J_F^{\ket{11} \rightarrow\ket{10}}+J_F^{\ket{10} \rightarrow\ket{11}}$.}.
Note that the control Hamiltonian is not diagonal in the natural computational basis. 
To obtain a proper representation for the Trotter exact derivatives we therefore numerically diagonalize $\H_n^c=u_n(\op b_1^\dagger + \op b_1)$. 
The eigenvectors are identical for all nonzero values of the control and we take $u_n=1$ for simplicity. 
Storing these eigenvectors as columns in the basis transformation operator $\op{\mathcal{R}}$, we perform the basis change,
\begin{align}
\H_n^c &\leftarrow \op{\mathcal{R}}^\dagger \H_n^c \op{\mathcal{R}}, &&\H^d \leftarrow \op{\mathcal{R}}^\dagger \H^d \op{\mathcal{R}}, \label{eq:basischange1}\\
\ket{\psio} &\leftarrow \op{\mathcal{R}}^\dagger \ket{\psio}, &&\hspace{-0.4cm}\ket{\psitgt} \leftarrow \op{\mathcal{R}}^\dagger \ket{\psitgt}. \label{eq:basischange2}
\end{align}
The results of optimizing this control-diagonalized problem are shown in Fig.~\ref{fig:transmon_example}. We find a nearly identical situation to Fig.~\ref{fig:2level_example}, except the gap between control-diagonal Trotter- and exact propagator gradient methods has significantly increased (note log-scale) due to the increased $D_\mathcal{H}$. 
Further, more iterations are generally needed and the first-order approximation ``falsely'' converges to even worse fidelities. This indicates that this problem is somewhat more challenging than that of Fig.~\ref{fig:2level_example}.

Performing optimizations at durations $0.5$, $0.75$, $1.25$ and $1.5$ times $T=2.83$ did not change the results outside of scaling the performance gap and the overall best attainable fidelity for the lower values. 
In contrast to the two-level problem, this problem was never so simple/easy that the approximate gradient could yield competitive results to the exact gradient in the exact landscape and we expect this to be representative of most control problems.

\subsubsection{Gradient Evaluation Time}
\begin{figure}[t]
	\includegraphics[]{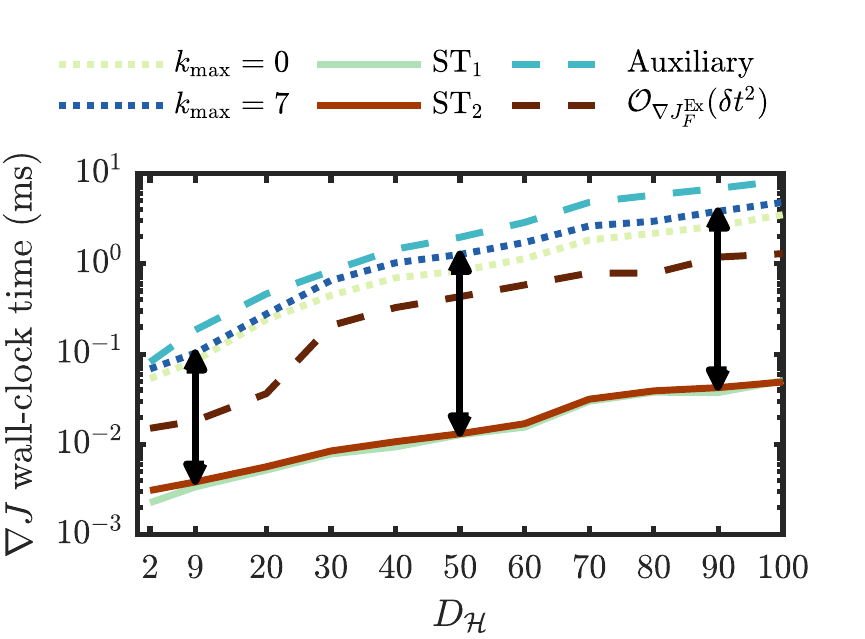}
	\caption{Gradient and recursive commutator tail Eq.~\eqref{eq:exactderiv_tail} median calculation time with negligible variance as a function of Hilbert space dimensionality. 
		The exponentially increasing performance gap between exact derivative methods is roughly indicated by the arrows.
	}
\label{fig:gradientwalltime}
\end{figure}

Next, we record the wall-clock time for calculating the various gradients as a function of $D_\mathcal{H}$, including $D_\mathcal{H}=2$ and $D_\mathcal{H}=9$ associated with Figs.~\ref{fig:2level_example}--\ref{fig:transmon_example}.
For each $D_\mathcal{H}$ we generate 10 random Hamiltonian matrices and controls of length $\nt = 400$. 
Figure~\ref{fig:gradientwalltime} shows the median wall-clock time consumption and these results reveal that the performance gap between the exact derivative methods is monotonically and exponentially increasing (note log-scale). 
This is not inherently surprising due to the exact propagation itself being trivially much slower. 
However, even when subtracting this contribution, calculating the recursive commutator tail in Eq.~\eqref{eq:exactderiv_tail}
to ensure gradient exactness exhibits a similar scaling with orders of magnitude in separation to the \textit{full} Trotter calculations. 
That is, the smaller gaps between the three upper methods also grow with $D_\mathcal{H}$. This trend was already visible in Figs.~\ref{fig:2level_example}--\ref{fig:transmon_example} by comparing relative wall-clock time distance between the dark and light blue trajectories.

A natural question is, then, where the exact derivatives for other approximate dynamical schemes such as the aforementioned Krylov-Lanczos, finite Taylor series, Crank-Nicolson, and Chebyshev would manifest in Fig.~\ref{fig:gradientwalltime}. Leaving a complete numerical study of this for future work, we nevertheless argue based on scaling properties in Sec.~\ref{sec:discussion} that Krylov-Lanczos, perhaps the most prominent general alternative, would lie somewhere in the shown performance gap for $D_\mathcal{H} \gg 1$ and above otherwise. 
We have performed similar mathematical analysis for the remaining mentioned schemes but consider it beyond the scope of this paper to include it explicitly since they yield comparable or worse results.
In addition to $\dt$, most of these also depend on series truncation parameters similar to $k\st{max}$ in Eq.~\eqref{eq:exactderiv_tail}.

\section{Generalizability of Multiple Controls}
\label{sec:generalization}
Up to now our analyses assumed a single, generically parametrized control Hamiltonian $\Hc(t,u(t))$. 
When including more than one control, our approach depends on the computational feasibility of maintaining the diagonality criteria for the control Hamiltonians as follows.

We denote a set of $K$ controls and their corresponding control Hamiltonians by
\begin{align}
C(t)=\{\Hc_{(k)}(t,u_{(k)}(t))\}_{k=1}^K = \cup_{q=1}^{Q} C_q(t), \label{eq:C}
\end{align}
that we have sorted into $Q$ sets $C_q$ of mutually commuting elements with $K = \sum_q |C_q(t)|$. 
Let $\R_q(t)$ be the unitary basis change operator that simultaneously diagonalizes the elements of $C_q(t)$ from a chosen reference basis e.g. a ``natural'' or the computational. That is, $[\op X ,\op Y] = 0$ and $\R_q(t)^\dagger \op X \R_q(t)$ is diagonal for all $\op X,\op Y\in C_q(t)$. 

Upon temporal discretization, consider the time evolution operator at time index $n$ with e.g. the $\ut{ST_2}$ expansion of Eq.~\eqref{eq:STProp2},
\begin{align}
&\U_n\ut{Ex} = e^{ -{i \dt} \left(\Hd_n + \sum_{k=1}^{K} \Hc_{n,(k)} \right)}   \notag \\
\approx \;&\U_n\ut{ST_2} = e^{ -\frac{i \dt}{2} \left(\sum\limits_{k}  \Hc_{n,(k)}\right) }
e^{ -{i \dt} \Hd_n }
e^{ -\frac{i \dt}{2} \left(\sum\limits_{k}  \Hc_{n,(k)}\right)}, 
\end{align}
The sum can be grouped as $\sum_k = \sum_{q} \sum_{C_{n,q}}$ and the exponential can then be split into the sets of mutually commuting elements with further first-order Trotterizations as in Sec.~\ref{sec:SuzukiTrotter}, 
\begin{align}
&e^{ -\frac{i \dt}{2} \sum_{q} \left(\sum_{C_{n,q}}  \Hc_{n,(k)}\right) } \approx \prod_{q=1}^{Q} e^{ -\frac{i \dt}{2} \left(\sum_{C_{n,q}}  \Hc_{n,(k)}\right) } \notag \\
&=\prod_{q=1}^{Q} \R_{n,q} \left(e^{ -\frac{i \dt}{2} \left( \sum_{C_{n,q}}  \R_{n,q}^\dagger \Hc_{n,(k)} \R_{n,q}\right) }\right) \R_{n,q}^\dagger, \label{eq:multicontrol}
\end{align}
which is exact if $Q=1$ and of order $\mathcal{O} (\dt^2)$ otherwise, and we utilized that $e^{\op A} = \op B e^{\op B^\dagger \op A \op B} \op B^\dagger$ for any unitary $\op B$.  Here, each $\R_{n,q}^\dagger \Hc_{n,(k)} \R_{n,q}$ is by definition diagonal, and  Eq.~\eqref{eq:multicontrol} thus sequentially transforms into the respective control-diagonal bases where it is trivial to compute the exponentials and their exact derivatives. This is the same core idea as in the split-step Fourier transform \cite{von2008computational} for propagating real space wave functions. \\

For this to be an effective approach for $Q>1$, performing the basis change $\R_{n,q}$ must be significantly cheaper than the original dense exponentiation. 
The one exceptional instance where this condition is not met occurs when $\R_{n,q}$ depends on the control value and, simultaneously, no closed analytical solution to the transformations are known.
This implies that $\R_{n,q}$ must be obtained anew in each iteration by numerical diagonalization, which is as expensive as dense exponentiation.
Otherwise, $\R_{n,q}$ and products involving these need only be calculated maximally once and can be stored on the disk and be loaded into memory at runtime. 
Note in particular that the exceptional case can be categorically ruled out upon additionally including the pervasive assumption of bilinear controls, i.e.  $\Hc_{(k)}(t,u(t)) \rightarrow u_{(k)}(t) \Hc_{(k)}(t)$. As in the transmon example, this is because the control value is simply a scaling factor to the matrix diagonalization.

In certain $Q>1$ cases it may not possible to numerically diagonalize for $\R_{n,q}$ even once due to very large $D_\mathcal{H}$ for example in a many-body setting. 
One is then seemingly restricted to $Q=1$ which incidentally covers a broad range of rich and realistic problems: for example combined control over (i) individual- and similar site-site couplings in spin chains \cite{murphy2010communication} ($\sum_{k} u_{k}(t) \op\sigma_k^{i}$ and $\sum_{k}  u_{k}(t) \op\sigma_k^{i}\op\sigma_{k+1}^{i}$, respectively, for any $i=x,y,z$), or (ii) on-site interaction- and single-site potentials in the Bose-Hubbard model \cite{jensen2020achieving,doria2011optimal} ($u(t) \sum_k \op n_k(\op n_k - 1 )$ and $\sum_k u_k(t) \op n_k$, respectively). 
However, we may yet entertain the capability of treating $Q>1$. 
For example, a $Q=2$ spin-chain on the form $\H = u_1(t)\sum \op\sigma_{k}^x \op \sigma_{k+1}^{x} + u_2(t)\sum \op\sigma_{k}^z \op\sigma_{k+1}^{z}$ can be represented either in terms of $\op\sigma^{x}$ or $\op\sigma^{z}$ eigenstates which yield control diagonal representations for the respective terms. 
We can then proceed to optimize one of the mutually commuting sets while momentarily considering the rest a drift contribution so effectively $Q=1$ for a number of iterations. 
That is, by numerically constructing the bases, Hamiltonians, and states for each set individually and choosing either of the representations at the start of each iteration we circumvent the need for explicitly calculating $\R_{n,q}$ that transforms between them.
The controls $u_1(t)$ and $u_2(t)$ in the example may then sequentially optimized and in the optimization literature this is known coordinate descent \cite{nocedal2006numerical}.
\\

The principal computational cost associated with considering multiple controls in Eq.~\eqref{eq:C} is therefore not the number of controls $K$ itself, but the number of mutually commuting sets $Q \leq K$ they distribute into.
Importantly, the number of basis transformations scales only linearly with $Q-1$ and typical values are $Q=1,2$.
For the simplest case $Q=1$ and time-independent $\R = \R_{n,1}$ the full time evolution $\U(T;0)$ is further simplified since $\R^\dagger \R = 1$.
The two concrete problems studied in Sec.~\ref{sec:exactderivatives} fall into this category. For the two-level (transmon) system, the ``natural'' basis did (did not) diagonalize the control and thus $\R=1$ ($\R$ was numerically obtained). \\

The presented control-diagonal Trotter methodology's results and relative performance capacity therefore generalize well to more than one control.

\section{Discussion and Outlook}
\label{sec:discussion}

We discussed how and why accurate derivatives are central to achieve high fidelities and convergence rates in derivative-based methods for generic optimization tasks. 
Examining common choices of time evolution scheme, the exact- and two Suzuki-Trotter expanded propagators, showed how these can be interpreted and related in terms of optimization landscapes.
We found their resulting analytically exact derivatives to differ vastly in complexity: assuming a diagonal control Hamiltonian for the Trotterized landscapes, we circumvented a detrimental infinite series, and highlighted many additional attractive properties compared to the exact landscape
leading to the performance gap qualitatively illustrated in Fig.~\ref{fig:landscape}.
Additionally, only one of the Trotter gradients' analytical form coincides with that of the first-order approximation in the exact landscape.
This is quite exceptional, since the second Trotter gradient of equivalent complexity --- and those due to other, standard dynamical approximations with increased derivative complexities --- do not.
When balancing respective errors in the dynamics and in the derivative calculations, the latter is in a certain sense more important as the derivatives should ``match'' the landscape.
We demonstrated the main ideas by considering two problems of varying Hilbert space size and in both instances found the expected performance hierarchies. The control-diagonal Trotter exact derivatives lead to orders of magnitude increase in computational speed and high-fidelity results with zero error to machine precision. This trend was verified to be monotonic and exponentially growing 
due to separate numerical complexity differences in both dynamics and derivatives with the Hilbert space size.
Finally, it was shown that the control-diagonal Trotter methodology generalizes well to more than one control. 

\medskip

The immediate advantages of optimizing in the Trotterized landscapes over the exact landscape are twofold: (i) they are applicable to much larger systems, and (ii) their analytically exact control derivatives and thus search directions essential for optimization convergence 
are greatly simplified.
We assumed that the control Hamiltonian is diagonal and this can practically always be, and often automatically is, fulfilled. 
In particular, the main computational effort in the optimization lies in propagating the state --- i.e. the dynamics --- and is not subject to severely scaling bottlenecks like matrix exponentiation, recursive commutators, or diagonalization. We did not include the exact gradient obtained in the diagonalization paradigm \cite{dalgaard2020hessian} for our comparative studies. The principal cost is $\nt$ diagonalizations of size $D_\mathcal{H} \times D_\mathcal{H}$ in each iteration and it would thus exhibit a similar performance gap to the Trotter methods.

The only implicit requirement is that $\dt$ is small enough for the Trotterization to faithfully approximate the exact dynamics, or equivalently, the exact landscape in Fig.~\ref{fig:landscape} and Eq.~\eqref{eq:landscape}. 
We did not explicitly include this consideration in the discussion of the presented examples since it is not central or important to the overall methodological performance hierarchy.
Of course, to obtain meaningful results in practical application, it is essential to establish such an upper bound for the time resolution. 
Note, however, that the exact Trotter derivatives are irrespective of $\dt$ in terms of complexity. This allows for the use of, among others, effective use of \textit{homotopy} methods \cite{borzi2017formulation,jensen2020achieving} in $\dt$, that is, optimization with an increasingly finer time resolution. This can be used to significantly speed up initial iterations without loss of accuracy in the final dynamics which typically has an error of order $\mathcal O(\dt^2)$, because the (quasi-)continuous deformation of $\dt$ can be made arbitrarily small at the end. 
The exact propagator can nonetheless still be used if the Hilbert space is sufficiently low-dimensional such that direct exponentiation and series summation or alternatively auxiliary matrix exponentiation or diagonalization is feasible c.f. Fig.~\ref{fig:gradientwalltime}. However, even for the simplest possible nontrivial two-level LZ problem illustrated in Fig.~\ref{fig:2level_example}, this approach is seen to be much slower than the alternative, controllably approximate methods. This difference in computational feasibility increases monotonically and exponentially with the Hilbert space dimensionality as evidenced by Fig.~\ref{fig:gradientwalltime}.
The aggregate computational performance of our control-diagonal Trotter methodology provides a scaffolding for efficient derivative-based optimization of very high-dimensional many-body dynamics in the high-fidelity limit. 
We pursue this in parallel work \cite{jensen2020achieving} for a system far beyond exact diagonalization approaches, necessitating a matrix product state description. 
Below we expand on a few pertinent discussion points.

\medskip

\medskip

\textit{Krylov-Lanczos methods. ---}
Another common way of approximating the time evolution for extended applicability is through the use of Krylov-Lanczos subspace methods \cite{park1986unitary,hochbruck1997krylov, beerwerth2015krylov} 
where exponential operator applications are performed without explicit construction.
Notice that the auxiliary matrix method Eq.~\eqref{eq:auxiliary} can be adapted to this setting by multiplying from the right with $\large(\vec 0, \ket{\psi_n}\large)^T$, yielding two separate Krylov-Lanczos calculations, $\large(\frac{\partial\U_n}{\partial u_n} \ket{\psi_n}, \U_n \ket{\psi_n}\large)^T$, for each of the $K$ controls in Eq.~\eqref{eq:multicontrol}.
Here we briefly compare this approach with our Trotterized control-diagonal scheme for $K=1$. For example, numerically stepping forward in time $\ket{\psi_{n+1}} = \U_n \ket{\psi_n}$ with Krylov-Lanczos entails (i) iterative construction of $k \leq D_\mathcal{H}$ Lanczos vectors $q_i$ of dimension $D_\mathcal{H}$ each requiring a matrix-vector multiplication on the form $\H_n q_i$ as the most expensive operation, and (ii) matrix exponentiation of a  $k \times k$ matrix and at least another matrix multiplication. This turns out to be computationally efficient compared to exact propagation when $1 < k \ll D_\mathcal{H}$, where $k$ controls the approximation accuracy. 
The control-diagonal Trotter steps e.g. $\ket{\psi_{n+1}} = \U_{n}^{c/2} \U_n^d \U_{n}^{c/2} \ket{\psi_{n}}$ numerically naïvely require a total of 3 matrix multiplications, but the vectorized form Eq.~\eqref{eq:vectorized} reduces this significantly. Since $\U_{n}^{c/2}$ is diagonal, the exponentiation of each diagonal element can be efficiently stored in a vector, and a total of two element-wise vector-vector multiplications need to be performed for the control part. For the drift part, recall  $\U_n^d$ only needs to be calculated once and can be cached indefinitely, leaving only a matrix-vector multiplication with the same cost as constructing a \textit{single} Lanczos vector. Thus, simply constructing the Lanczos vectors (i) is more costly than performing the full Trotter step. 
Further, the Krylov-Lanczos procedure obfuscates the direct analytical dependence on the control $u_n$, disallowing a straightforward analytical derivative calculation. 
As opposed to the control-diagonal Trotterization, this leads to a ``mismatch'' between the optimization landscape and the derivative calculations unless $k$ is large enough since their exactness are both linked to an approximation parameter. 
Nevertheless, Krylov propagation is much preferable to exact propagation even for moderate values of $D_\mathcal{H}$ and we expect that further studies would place it somewhere inside the performance gaps in Figs.~\ref{fig:2level_example}--\ref{fig:gradientwalltime}.


\medskip
\textit{Robustness. ---}
Optimal controls extracted from open-loop methodologies may be sensitive to variations and uncertainties in the underlying physical model.
Let $x$ be a physical and possibly time dependent quantity related to the control or any of the uncontrolled system parameters.
If $x$ was modeled to have value $x\st{theory}$ but in actual experimental implementation has value $x\st{experiment}$ then the fidelity is likely to degrade as a result, $F\st{experiment} < F\st{theory}$. 
These modeling errors could originate from many sources, for example imperfect equipment fabrication, drifting or fluctuating noisy signals, or nondeterministic run-to-run system preparation. 

It is, however, possible to account for such errors by including these uncertainties through, e.g., \textit{ensemble optimization} \cite{goerz2014robustness,sorensen2020optimization}.
The cost function is then taken as an ensemble average $\bar J = \sum_{l=1}^{L} J[x_l]$, possibly weighted, over $L$ realizations of the physical system. In each realization, the uncertain parameter is taken to be $x=x_l$  which could be randomly or regularly sampled from a suitable model distribution, for example Gaussian or bounded uniform. 
If the noise on $x$ is adequately characterized, the optimal controls achieved by minimizing $\bar J$ will have built-in robustness to the parameter fluctuations. 

Although this idea is simple and straightforward to implement, each iteration now requires the calculation of $L$ gradients since $\frac{\partial \bar J}{\partial u_n} = \sum_{l=1}^{L} \frac{\partial J[x_l]}{\partial u_n}$.
This places an increased emphasis on both gradient computation speed and exactness.
Reaching a certain number of iterations increases the wall-clock time by roughly a factor of $L$. 
Additionally, derivative errors are compounded much more severely between iterations. 
For example, Hessian approximation errors for quasi-Newton methods would lead to false convergence at roughly $L$ times the normal rate or equivalently at $1/L$ times the normal iterations. 

As example, suppose one of the physical parameters associated with Figs.~\ref{fig:2level_example}--\ref{fig:transmon_example} 
was uncertain and optimized with $L=10$ ensemble members. 
The $k\st{max}=0$ approximation in Figs.~\ref{fig:2level_example}--\ref{fig:transmon_example} stagnated at $<50$ iterations and would now stagnate at $<10$. The exact gradient wall-clock time performance gap between the Suzuki-Trotter and exact propagator methods would increase by another order of magnitude in absolute time since all iterations take $L=10$ longer.

The herein presented control-diagonal Trotter methods are thus not only prospectively useful for handling larger Hilbert spaces but also for more efficiently incorporating robustness. 
%
%
%
%
%
%

\medskip
\textit{Hessian. ---}
The exact Hessian has strong theoretical properties as discussed in Sec.~\ref{sec:exactderivatives}, and although our new calculations of the exact Hessian have been verified numerically, we did not yet perform comparative studies and leave this to future work. Nevertheless, we have found in a parallel, similar work that a novel calculation of the exact Hessian within the diagonalization paradigm \cite{dalgaard2020hessian} outperforms a gradient-only quasi-Newton approach in terms of statistics and best results in certain domains. This suggests similar possibilities in the present case.

\medskip
\textit{Discretize-then-optimize. ---}
Lastly, we point out that the results in this paper followed a discretize-then-optimize (time discretization before ordinary vector derivatives of a cost function) rather than optimize-then-discretize \cite{sorensen2019qengine} (time discretization after continuous G\^ateaux derivatives of a cost functional) approach. 
Since these approaches do not in general necessarily yield the same derivative expressions, the former approach is preferable because it specifically takes into account the chosen propagation scheme implementation and the derivatives ``match'' the landscape/dynamics which has been a main point throughout this manuscript 
\footnote{Similarly, applying of Krylov subspace methods for the time evolution while using exact gradients for the exact propagator constitutes another potential mismatch between the derivatives and landscape/dynamics.}. 
It is therefore quite fortuitous that e.g. (i) the exact propagator gradient $\nabla J_\F\ut{Ex}$ calculated by the optimize-then-discretize approach \cite{sorensen2019qengine} yields exactly the same expression as $\nabla J_\F\ut{ST_1}$, and (ii) $\U\ut{ST_1}$ is a standard propagator for some systems, e.g. for wave functions in real space \cite{borzi2017formulation,von2008computational,hohenester2007optimal}. 
The combined effect is that the sought-after gradient exactness is obtained by virtue of standard methods alone in these situations. 
The same would not be true if either $\U\ut{Ex}$ or $\U\ut{ST_2}$ was used cf. Eqs.~\eqref{eq:gradients}. 
%


\begin{acknowledgements}
We thank I. Kuprov for useful discussions and C.A. Weidner and M. Dalgaard for feedback.
This work was funded by the ERC, H2020 grant 639560 (MECTRL), and the John Templeton and Carlsberg Foundations.
\end{acknowledgements}

\appendix

%

\section{Derivation of Exact Gradients and Hessians}
\label{app:derivatives}
Here we present the calculations leading to the exact gradient and Hessian expressions for one of the exact exponentiation propagators and the two Trotterized propagators with a diagonal control Hamiltonian defined in Eqs.~\eqref{eq:propagators}. We also define the regularization cost functionals and likewise calculate their derivatives after discretization. 
Emphasis is put on thoroughness of the steps, and relevant equations for the derivations are restated for convenience where applicable so as to be self-contained. \\

We assume that the control $u(t)$ is discretized on a regularly spaced time grid $t \in [t_1,t_2,\dots,t_{\nt}] = [0, \dt, \dots, T]$. Recall
$\ket{\psi_{n+1}} = \U_n\ket{\psi_n}$ with $\ket{\psi_1}= \ket{\psio}$ and $\ket{\psi_{\nt}}= \ket{\psi(T)}$, and $\H(u_n) = \H_n = \Hd_n + \Hc_n$ where $\Hc_n$ ($\Hd_n$) is the control (drift) Hamiltonian. Define the auxilliary state $\ket{\chi_{\nt}} = \ket{\psitgt}$ with $\ket{\chi_{n}} = \U_{n}^\dagger \ket{\chi_{n+1}}$. 
The derivatives given in Eqs.~\eqref{eq:derivativeequations} are to be evaluated for the propagators $\U\ut{Ex} = \U\ut{Ex_2}$, $\U\ut{ST_1}$, and $\U\ut{ST_2}$ given in Eqs.~\eqref{eq:propagators}.

As mentioned in the main text, the following results are trivially extended to situations other than pure state transfer: the central calculations of Eq.~\eqref{eq:gradhess} are the same. The primary differences lie in how these enter a given cost functional and if the cached objects are states or either unitary- or density matrices. 


\subsection{Derivatives for Exact Propagator}
The exact exponentiation propagator has the form $\U_n\ut{Ex} = \exp(-i\op H(u_n) \delta t)$ and the derivative overlap reads
\begin{align}
\frac{\partial o}{\partial u_n} &= \Braket{\chi_{\nt}  |  \frac{\partial}{\partial u_n}  \left(\U_{\nt-1}\ut{Ex}\dots \U_n\ut{Ex}\dots \U_1\ut{Ex}\right)  | \psi_1}  \notag\\ 
&= \Braket{\chi_{n+1}  |  \frac{\partial \U_n\ut{Ex}}{\partial u_n} | \psi_n}. \label{eq:overlapderivative}
\end{align}
The task is then to calculate $ \frac{\partial \U_n\ut{Ex}}{\partial u_n}$ and be careful with ordering. 
The $\ut{Ex}$ superscript is omitted for brevity in most of the steps below.
We expand the exponential as 
\begin{align}
\frac{\partial \U_n}{\partial u_n} &= \frac{\partial}{\partial u_n} \left( e^{-i\H_n \dt} \right) = \sum_{p=0}^{\infty} \frac{(-i \delta t)^p}{p!} \frac{\partial}{\partial u_n} {\bigg( \H_n^p \bigg)} \notag\\
&= \sum_{p=1}^{\infty} \frac{(-i \delta t)^p}{p!}  \sum_{q=0}^{p-1} \H_n^q \left(\frac{\partial \H_n}{\partial u_n}\right) \H^{p-q-1},
\end{align}
Define for momentary simplicity $A\equiv -i \H_n \dt$ and $B\equiv -i \frac{\partial \H_n}{\partial u_n} \dt$. Then it can be shown that 
\begin{align}
\frac{\partial \U_n}{\partial u_n} \overset{}{=}  \sum_{p=0}^{\infty} \sum_{q=0}^{\infty} \frac{A^p B A^q}{(p + q + 1)!}  .
\end{align}
Using now the following relations for the Beta and Gamma functions \cite{boas2006mathematical}:
\begin{align} 
\Gamma(a) &= (n-1)!  \hspace{1cm} (\mathrm{for}\; a\in \mathbb Z^+),\\
\beta(a,b) &=  \int_{0}^1 (1-\alpha)^{a-1} \alpha^{b-1} \d \alpha = \frac{\Gamma(a) \Gamma(b)}{\Gamma(a+b)}  \notag\\
&= \frac{(a-1)! (b-1)!}{(a+b-1)!} = \frac{p! q!}{(p+q+1)!},
\end{align}
and taking $a=p+1, \;b = q+1 $ we obtain
\begin{align}
\frac{1}{(p+q+1)!} =  \frac{1}{p!q!} \int_{0}^1 (1-\alpha)^{p}\alpha^{q} \d \alpha.
\end{align}
Inserting this, and initially pulling out the integral we obtain 
\begin{align}
\frac{\partial \U_n}{\partial u_n} &= 
\sum_{p=0}^{\infty} \sum_{q=0}^{\infty} \frac{A^p B A^q}{p!q!}   \int_{0}^1 \alpha^{p}(1-\alpha)^{q} \d \alpha \notag \\
&=  \int_{0}^1 \left(\sum_{p=0}^{\infty}\frac{((1-\alpha) A)^p}{p!}\right) B \sum_{q=0}^{\infty}  \frac{(\alpha A)^q}{q!}  \d \alpha \notag \\
&=  \int_0^1 e^{(1-\alpha) A} B e^{\alpha A} \d \alpha =  e^{A} \int_0^1 e^{-\alpha A} B e^{\alpha A} \d \alpha \notag  \\
&=  \U_n \int_0^1 e^{(i \alpha  \dt) \H_n} \bigg(-i\dt \frac{\partial \H_n}{\partial u_n}\bigg) e^{-(i \alpha \dt) \H_n} \d \alpha.
\end{align} 
The integrand can be evaluated by defining the recursive commutator in Eq.~\eqref{eq:recursivecomm} with base case $[c_x X,c_y Y]_0 = c_y Y$ and using Baker-Campbell-Hausdorff relations \cite{sakurai2017modern}
\begin{align}
[c_x X,c_yY]_k &= [c_xX,[c_x X,c_y Y]_{k-1}] = c_x^{k} c_y [X,Y]_k, \\
e^{c_x X}Y e^{-c_x X} &= \sum_{k=0}^{\infty} \frac{[c_x X,c_y Y]_k}{k!} = \sum_{k=0}^{\infty}  \frac{c_x^k c_y}{k!}[X,Y]_k,
\end{align}
and by evaluating these with scalars $c_x = i\alpha \dt$ and $c_y = -i\dt$, one obtains
\begin{align}
\frac{\partial \U_n}{\partial u_n} &= \U_n \int_0^1 \left( \sum_{k=0}^{\infty} \frac{(i\alpha \dt )^k (-i\dt)}{k!} [\H_n,\frac{\partial \H_n}{\partial u_n}]_k \d \alpha \right) \notag \\
&= \U_n   \sum_{k=0}^{\infty} (-i \dt) \frac{i^k \dt^{k} }{k!} [\H_n,\frac{\partial \H_n}{\partial u_n}]_k \left( \int_0^1 \alpha^k\d \alpha \right) \notag \\
&= \U_n   (-i\dt)  \sum_{k=0}^{\infty}  \frac{i^k \dt^{k} }{(k+1)!} [\H_n,\frac{\partial \H_n^{}}{\partial u_n} ]_k. \label{eq:dUdu_app}
\end{align}
Substituting this into Eq.~\eqref{eq:overlapderivative}, then inserting the resulting expression into Eq.~\eqref{eq:derivativeequations}, and using $\frac{\partial \H_n}{\partial u_n} = \frac{\partial \H_n^{c}}{\partial u_n}$ gives 
\begin{align}
&\frac{\partial J_\F \ut{Ex}}{\partial u_n} =   - \Re\left(\overlap^* \Braket{\chi_{n+1}  |  \frac{\partial \U_n\ut{Ex}}{\partial u_n} | \psi_n}\right)  \notag \\
&= \Re\left( i\overlap^* \Braket{\chi_{n} | \left(  \sum_{k=0}^{\infty}  \frac{i^k \dt^{k} }{(k+1)!} [\H_n, \frac{\partial \H_n^{c}}{\partial u_n}]_k \right)| \psi_{n}  }\right) \dt \notag \\
&= \Re \left(i \overlap^*  \Braket{\chi_n |  \frac{\partial \H_n^{c}}{\partial u_n} | \psi_{n}} \right)  \dt  + \mathcal O_{\nabla J_F\ut{Ex}}(\dt^2)
\end{align}
which is the expression stated in Eqs.~\eqref{eq:gradJexact} and \eqref{eq:exactderiv_tail}.

With the gradient at hand, the Hessian calculation only needs additional evaluation of the second derivatives of $\overlap$, 
\begin{subequations}
\begin{align} 
n>m: \hspace{0.45cm} &\notag \\
\frac{\partial^2 o}{\partial u_n \partial u_m}  &= \braket{\chi_{\nt}|\U_{\nt-1} \dots \frac{\partial  \U_n}{\partial u_n} \dots \frac{\partial  \U_m}{\partial u_m} \dots \U_1 |\psi_1} \notag\\
&= \braket{\chi_{n+1}|\frac{\partial  \U_n}{\partial u_n} \left(\prod_{j=m+1}^{n-1}\U_j \right) \frac{\partial  \U_m}{\partial u_m} |\psi_{m} },\\
n=m: \hspace{0.55cm}  & \notag \\
\quad \frac{\partial^2 o}{\partial u_n \partial u_m}  &= \braket{\chi_{\nt}|\U_{\nt-1} \dots \frac{\partial^2  \U_n}{\partial u_n^2} \dots\U_1 |\psi_1} \notag \\
&= \braket{\chi_{n+1}|\frac{\partial^2  \U_n}{\partial u_n^2} |\psi_{n} } .
\end{align}  
\end{subequations}
The case $m>n$ is the same as $n>m$ with indices $n \rightleftarrows m$ and we thus need only calculate one of the cases due to this symmetry. 
Inserting these expressions in Eqs.~\eqref{eq:derivativeequations} 
we obtain the exact Hessian elements $n\geq m$ without loss of generality for the exact propagator $\U_n = \U_n\ut{Ex}$
\begin{align}
&\frac{\partial^2 J_\F\ut{Ex}}{\partial u_n \partial u_m}  
= - \Re\left(\!\Big(\!\!\braket{\chi_{m+1} | \frac{\partial  \U_m}{\partial u_m} |\psi_m}\!\!\Big)^{\!\!*}\! \braket{\chi_{n+1}|\frac{\partial  \U_n}{\partial u_n} |\psi_n }\right) \notag\\
&-\Re\left( o^* \braket{\chi_{n+1}|\frac{\partial  \U_n}{\partial u_n} \Big(\prod_{j=m+1}^{n-1}\!\!\!\U_j \Big) \frac{\partial  \U_m}{\partial u_m} |\psi_{m}}  \right)\! (1-\delta_{n,m}) \notag\\
&-\Re\left(o^*  \braket{\chi_{n+1}|\frac{\partial^2  \U_n}{\partial u_n^2} |\psi_{n} }  \right) \delta_{n,m},
\label{eq:exactHessian}
\end{align}
where $\frac{\partial  \U_n}{\partial u_n}$ is given by Eq.~\eqref{eq:dUdu_app}.
Note the third and second term appear only on the diagonal and off-diagonal, respectively.
In optimization contexts the propagator gradient is also always computed and those elements can thus be reused here in practical applications. 
Evaluating $\frac{\partial^2  \U_n}{\partial u_n^2}$ is straightforward and gives,
\begin{align}
\frac{\partial^2  \U_n}{\partial u_n^2} 
&=   \U_n \bigg\{ \bigg(   -i\dt \sum_{k=0}^{\infty}  \frac{i^k \dt^{k} }{(k+1)!} [\H_n, \frac{\partial \H_n^{c}}{\partial u_n}]_k\bigg)^2 \notag\\
   -& i\dt  \sum_{k=0}^{\infty}  \frac{i^k \dt^{k} }{(k+1)!}  \frac{\partial }{\partial u_n} \left([\H_n, \frac{\partial \H_n^{c}}{\partial u_n}]_k \right)\bigg\}, \label{eq:ddUddu}
\end{align}
but the recursive commutator derivative is cumbersome
\begin{align}
[\H_n,\H_n']'_{k} &=  [\H_n,[\H_n,\H_n']'_{k-1}]-[\H_n',[\H_n,\H_n']_{k-1}] \\
[\H_n,\H_n']'_{0} &= \H_n'' ,\notag \\
[\H_n,\H_n']'_{1} &= [\H_n,\H_n''],  \notag\\ 
[\H_n,\H_n']'_{2} &= [\H_n,[\H_n,\H_n'']]  + [\H_n',[\H_n,\H_n']] ,\notag\\
&\hspace{0.2cm}\vdots \notag
\end{align}
where we explicitly evaluated the first few terms and $\H_n' \equiv  {\partial \H_n}/{\partial u_n}$. Note that the exact derivatives both entail an infinite summation or to machine precision in finite arithmetic.

\subsection{Derivatives for Trotterized Propagators}
We consider now in turn the Suzuki-Trotter expansions $\,\U_n\ut{ST_1} = \U_{n+1}^{c/2} \U_n^d \U_{n}^{c/2}$ and $\,\U_n\ut{ST_2} = \U_{n}^{c/2} \U_n^d \U_{n}^{c/2}$ where $\U_n^{c/2} \equiv \exp({-i\H_{n}^c \dt/2})$ and $\U_n^d$ is given by Eq.~\eqref{eq:driftexponential}.

\subsubsection{Derivatives of $\U_n\ut{ST_1}$}
For the Suzuki-Trotter expansion $\U_n\ut{ST_1} = \U_{n+1}^{c/2} \U_n^d \U_{n}^{c/2}$, the $u_n$ control dependence is distributed among $n$ and $n-1$ (except at the end points $n=1,N$), yielding 
\begin{subequations}
\begin{align}
\frac{\partial o}{\partial u_n} &= 
\Braket{\chi_{\nt}|\frac{\partial}{\partial u_n} \left(\U_{\nt-1}\ut{ST_1}\dots \U_{n}\ut{ST_1} \U_{n-1}\ut{ST_1}  \dots \U_{1}\ut{ST_1}\right)|\psi_{1} } \notag \\
&=\Braket{\chi_{n+1}|\frac{\partial}{\partial u_n} \left(\U_{n}\ut{ST_1} \U_{n-1}\ut{ST_1}\right)|\psi_{n-1} }. \label{eq:exactGradGradST_deriv}
\end{align}
\end{subequations}
Additionally, assume that the control Hamiltonian is diagonal. We use Eq.~\eqref{eq:dUdu_app} to take the derivative of $\left(\U_{n}\ut{ST_1} \U_{n-1}\ut{ST_1}\right)$,
\begin{align}
&\frac{\partial}{\partial u_n}\left(\U_{n}\ut{ST_1} \U_{n-1}\ut{ST_1}\right) \notag\\
&=  \frac{\partial}{\partial u_n}\bigg(\left(\U_{n+1}^{c/2} \U_{n}^d \U_{n}^{c/2} \notag \right) \left(\U_n^{c/2} \U_{n-1}^d \U_{n-1}^{c/2} \right) \bigg) \\ &= 
\left(\U_{n+1}^{c/2} \U_{n}^d\right) \frac{\partial \U_{n}^{c}}{\partial u_n}  \left(\U_{n-1}^d \U_{n-1}^{c/2} \right) \notag\\
&=\left(\U_{n+1}^{c/2} \U_{n}^d\right) \left\{ \U_{n}^{c} \left(-i\dt \frac{\partial \H_n^{c}}{\partial u_n}\right) \right\} \left(\U_{n-1}^d \U_{n-1}^{c/2} \right) \notag \\
&= 
(-i\dt) \cdot  \left(\U_{n+1}^{c/2} \U_{n}^d \U_{n}^{c/2}\right)  \frac{\partial \H_n^{c}}{\partial u_n} \left(\U_{n}^{c/2}\U_{n-1}^d \U_{n-1}^{c/2} \right) \notag\\
& =(-i\dt) \cdot \U_{n}\ut{ST_1}  \frac{\partial \H_n^{c}}{\partial u_n} \U_{n-1}\ut{ST_1}.
\label{eq:dUUdu}
\end{align}
Here we also used the fact that two diagonal matrices always commute, first to evaluate the recursive commutator $[\H_n^c,\frac{\partial \H_n^{c}}{\partial u_n}]_k = \frac{\partial \H_n^{c}}{\partial u_n} \cdot \delta_{0,k}$ from Eq.~\eqref{eq:dUdu_app}, and second to recombine the initial propagators since $[\frac{\partial \H_n^{c}}{\partial u_n}, \U_{n}^{c/2}] = 0$.
Inserting this into Eqs.~\eqref{eq:exactGradGradST_deriv} and \eqref{eq:derivativeequations} yields the exact gradient stated in Eq.~\eqref{eq:exactGradientST},
\begin{align}
\frac{\partial J_\F\ut{ST_1}}{\partial u_n} &=  \Re
\left(i \overlap^*\Braket{\chi_{n} | \frac{\partial H_n^{c}}{\partial u_n} | \psi_{n}  } \right)\dt, 
\end{align}
with an additional factor $1/2$ at the end points ($n=1,\nt$). Apart from these, this is identical in structure to the exact propagator gradient Eq.~\eqref{eq:gradJexact} when in the first-order approximation $k\st{max}=0$, i.e. discarding the otherwise expensive $\mathcal O(\dt^2)$ tail.

%
The second derivatives to be calculated for the Hessian are 
\begin{subequations}
\begin{align}
n>m: \hspace{0.45cm} \notag \\
 \frac{\partial^2 o}{\partial u_n \partial u_m} \notag
 = &\bra{\chi_{n+1}} \frac{\partial}{\partial u_n}  \left(\U_{n}\ut{ST_1} \U_{n-1}\ut{ST_1}\right)  \\
 &\hspace{-1cm}\left(\prod_{j=m+1}^{n-2} \U_j \right)\frac{\partial}{\partial u_m}  \left(\U_{m}\ut{ST_1} \U_{m-1}\ut{ST_1}\right)\ket{\psi_{m-1} } \label{eq:n>m}, \\
n=m: \hspace{0.45cm} \notag \\ 
 \frac{\partial^2 o}{\partial u_n \partial u_m}&= \braket{\chi_{n+1}|\frac{\partial^2}{\partial u_n^2} \left(\U_{n}\ut{ST_1} \U_{n-1}\ut{ST_1}\right)|\psi_{n-1} }.  \label{eq:n=m}
\end{align}
\end{subequations}
For $n>m$  we only need the first derivative given in Eq.~\eqref{eq:dUUdu}. 
For $n=m$  we take the second derivative of $(\U_{n}\ut{ST_1} \U_{n-1}\ut{ST_1})$ using Eq.~\eqref{eq:dUUdu} and $[\frac{\partial \H_n^{c}}{\partial u_n}, \U_{n}^{c/2}] = 0$ 
\begin{align}
&\frac{\partial^2}{\partial u_n^2} \left(\U_{n}\ut{ST_1} \U_{n-1}\ut{ST_1}\right)  =  (-i\dt) \notag
\frac{\partial}{\partial u_n}\bigg(\U_{n}\ut{ST_1}  \frac{\partial \H_n^{c}}{\partial u_n}  \U_{n-1}\ut{ST_1}\bigg) \\
&= (-i\dt) 
\U_{n+1}^{c/2} \U_{n}^{d}
\frac{\partial}{\partial u_n}\bigg(\U_{n}^{c}  \frac{\partial \H_n^{c}}{\partial u_n} \bigg)   \U_{n-1}^{d} \U_{n-1}^{c/2} \notag \\
&=  (-i\dt) \U_{n+1}^{c/2} \U_{n}^{d} 
\bigg(\U_{n}^{c} \Big(-i\dt \frac{\partial \H_n^{c}}{\partial u_n}\Big)\frac{\partial \H_n^{c}}{\partial u_n}  \notag\\
&\hspace{2.5cm} + \U_{n}^{c} \frac{\partial^2 \H_n^{c}}{\partial u_n^2}\bigg)   \U_{n-1}^{d} \U_{n-1}^{c/2} \notag\\
&=  (-i\dt) \U_{n+1}^{c/2} \U_{n}^{d}   \U_{n}^{c/2}  
\bigg(\frac{\partial^2 \H_n^{c}}{\partial u_n^2} \notag \\
&\hspace{2.5cm}-i\dt \Big(\frac{\partial \H_n^{c}}{\partial u_n}\Big)^2 \bigg)   \U_{n}^{c/2} \U_{n-1}^{d} \U_{n-1}^{c/2} \notag
\\
&=
(-i\dt) \cdot  \U_{n}\ut{ST_1}  \left(
\frac{\partial^2 \H_n^{c}}{\partial u_n^2}
-i\dt   \Big(\frac{\partial \H_n^{c}}{\partial u_n}\Big)^2  
\right) \U_{n-1}\ut{ST_1}.
\label{eq:dUUduu}
\end{align}
Inserting into Eqs.~\eqref{eq:derivativeequations}, the exact Hessian elements $n\geq m$ for this Trotterization scheme are therefore
\begin{align}
&\frac{\partial^2 J_\F\ut{ST_1}}{\partial u_n \partial u_m} = -\Re\bigg(
\braket{ \psi_{m} | \frac{\partial \H_m^{c}}{\partial u_m}|\chi_{m}} \braket{\chi_{n} | \frac{\partial \H_n^{c}}{\partial u_n} | \psi_{n}} \bigg) \dt^2 \notag \\
&+\Re \bigg(\overlap^*
\braket{\chi_{n}| \frac{\partial \H_n^{c}}{\partial u_n}   \Big(\!\!\prod_{\substack{j=m}}^{n-1} \!\U_j\ut{ST_1}\!\Big) \frac{\partial \H_m^{c}}{\partial u_m} |\psi_{m} }  
\bigg) (1-\delta_{n,m}) \dt^2\notag \\
&+\Re \bigg( i \overlap^* \braket{\chi_{n}| \bigg(
	\frac{\partial^2 \H_n^{c}}{\partial u_n^2}
	-i\dt   \bigg(\frac{\partial \H_n^{c}}{\partial u_n}\bigg)^{\!\!2}  
	\bigg) |\psi_{n} } \bigg)  \delta_{n,m} \dt, \label{eq:exactHessianST}
\end{align}
Derivatives of the end points corresponding to the outer ``rim'' of the Hessian matrix carry an additional factor 1/2 each, for a total of $1/4$ in the corners and $1/2$ on the edges.
Note the third and second term appear only on the diagonal and off-diagonal, respectively.
As with the gradient, the Hessian is similarly identical to Eq.~\eqref{eq:exactHessian} when retaining only the $k=k\st{max} = 0$ term. 
As an implementation detail, note that the propagated states and operator-state products from \eqref{eq:exactGradientST} may be reused here. The second term is the most costly to evaluate because of additional state propagations. The order of evaluation should be done row-by-row to further increase reusability of computations.

\subsubsection{Derivatives of $\U_n\ut{ST_2}$}
The Suzuki-Trotter expansion reads $\U_n\ut{ST_2} = \U_{n}^{c/2} \U_n^d \U_{n}^{c/2}$, and the overlap derivative reads 
	\begin{align}
	\frac{\partial o}{\partial u_n} &= 
	 \frac{\partial}{\partial u_n}  \left(\U\ut{ST_2}_{\nt-1}\dots \U_n\ut{ST_2}\dots \U_1\ut{ST_2}\right) \notag \\
	&=\Braket{\chi_{n+1}|\frac{\partial \U_{n}\ut{ST_2}}{\partial u_n} |\psi_{n} }.  \label{eq:doduST2}
	\end{align}
Invoking Eq.~\eqref{eq:dUdu_app} for $\U_{n}^{c/2}$ we find 
\begin{align}
\frac{\partial \U_{n}\ut{ST_2}}{\partial u_n}  &= \frac{\partial \U_{n}^{c/2}}{\partial u_n} \U_n^d \U_{n}^{c/2}  + \U_{n}^{c/2}  \U_n^d \frac{\partial \U_{n}^{c/2}}{\partial u_n} \notag \\
&=-\frac{i\dt}{2} \left(\U_{n}^{c/2} \frac{\partial \H_n^{c}}{\partial u_n}  \U_n^d \U_{n}^{c/2} +\U_{n}^{c/2}  \U_n^d \U_{n}^{c/2} \frac{\partial \H_n^{c}}{\partial u_n}\right) \notag  \\
&=-\frac{i\dt}{2} \left(\frac{\partial \H_n^{c}}{\partial u_n} \U_{n}\ut{ST_2}  + \U_{n}\ut{ST_2} \frac{\partial \H_n^{c}}{\partial u_n}\right).
\end{align}
Here we also used that two diagonal matrices always commute, first to evaluate the recursive commutator $[\H_n^c,\frac{\partial \H_n^{c}}{\partial u_n}]_k = \frac{\partial \H_n^{c}}{\partial u_n}\cdot \delta_{0,k}$ from Eq.~\eqref{eq:dUdu_app}, and second to recombine the initial propagators since $[\frac{\partial \H_n^{c}}{\partial u_n}, \U_{n}^{c/2}] = 0$.
Substituting back into Eqs.~\eqref{eq:doduST2},\eqref{eq:derivativeequations} we find 
\begin{align}
&\frac{\partial o}{\partial u_n} =-\frac{i\dt}{2} \Braket{\chi_{n+1}|\left(\frac{\partial \H_n^{c}}{\partial u_n} \U_{n}\ut{ST_2}  + \U_{n}\ut{ST_2} \frac{\partial \H_n^{c}}{\partial u_n}\right) |\psi_n} \notag \\
&= -\frac{i \dt}{2} \bigg\{\Braket{\chi_{n+1}|\frac{\partial \H_n^{c}}{\partial u_n}| \psi_{n+1}} \notag 
+ \Braket{\chi_{n} |  \frac{\partial \H_n^{c}}{\partial u_n}| \psi_{n}} \bigg\} \\
& = -\frac{i \dt}{2} \sum_{p=n}^{n+1} \Braket{\chi_{p} | \frac{\partial \H_n^{c}}{\partial u_n} | \psi_{p}} \label{eq:dodunST2}\\
&\Rightarrow \frac{\partial J_\F\ut{ST_2}}{\partial u_n} =\Re \left(\frac{i \overlap^*}{2} \sum_{p=n}^{n+1} \Braket{\chi_{p} |  \frac{\partial H_n^{c}}{\partial u_n} | \psi_{p}} \right)\dt, 
\end{align}
for all $n=1,\dots,\nt-1$ which is the expression in Eq.~\eqref{eq:exactGradientST2}.

After some lines of calculation, the second derivatives of $o$ for the Hessian evaluate to
\begin{subequations}
	\begin{align}
	n>m: \hspace{0.45cm} \notag \\
	\frac{\partial^2 o}{\partial u_n \partial u_m} =\notag\\
	 &\hspace{-1.5cm} -\frac{\dt^2}{4}\sum_{p=n}^{n+1}\sum_{q=m}^{m+1} \Braket{\chi_p|\frac{\partial \H_n^{c}}{\partial u_n} \left(\prod_{j=q}^{p-1} \U_n\ut{ST_2}\right)\frac{\partial \H_m^{c}}{\partial u_m}|\psi_q}, \label{eq:dodunmST2}\\
	n=m: \hspace{0.45cm} \notag \\ 
	& \hspace{-1.7cm}	\frac{\partial^2 o}{\partial u_n \partial u_m}= \notag \\
	&\hspace{-1.5cm} -\frac{i\dt}{2}
	\bigg( \sum_{j=n}^{n+1} \Braket{\chi_j| \left(\frac{\partial^2 \H_n^{c}}{\partial u_n^2} - \frac{i\dt}{2} \Big(\frac{\partial \H_n^{c}}{\partial u_n}\Big)^2\right) | \psi_j}  \notag 
	\\
	&\hspace{0cm}-i \dt \Braket{\chi_{n+1}| \frac{\partial \H_n^{c}}{\partial u_n} \U_n\ut{ST_2} \frac{\partial \H_n^{c}}{\partial u_n}|\psi_n}
	\bigg),\label{eq:dodunnST2}
	\end{align}
\end{subequations}
Substituting Eqs.~\eqref{eq:dodunST2}, \eqref{eq:dodunmST2}, \eqref{eq:dodunnST2} into Eq.~\eqref{eq:derivativeequations} yields the final result,
\begin{align}
&\frac{\partial^2 J_\F\ut{ST_2}}{\partial u_n \partial u_m} = \Re\bigg\{\notag\\
&+\frac{1}{4} \sum_{p=n}^{n+1}\sum_{q=m}^{m+1} \braket{\psi_{q} | \frac{\partial \H_m^{c}}{\partial u_m} |\chi_{q} }  \braket{\chi_{p} |\frac{\partial \H_n^{c}}{\partial u_n} | \psi_{p}} \dt^2 \notag\\
&+\frac{o^*}{4}\sum_{p=n}^{n+1}\sum_{q=m}^{m+1} \braket{\chi_p|\frac{\partial \H_n^{c}}{\partial u_n} \prod_{j=q}^{p-1} \U_n\ut{ST_2} \frac{\partial \H_m^{c}}{\partial u_m}|\psi_q}(1-\delta_{n,m}) \dt^2 \notag\\
&+\frac{i o^*}{2}
\bigg( \sum_{j=n}^{n+1} \braket{\chi_j| \left(\frac{\partial^2 \H_n^{c}}{\partial u_n^2}- \frac{i\dt}{2} \Big(\frac{\partial \H_n^{c}}{\partial u_n} \Big)^2\right) | \psi_j}  \notag
\\
&\hspace{0.95cm}-i \dt \braket{\chi_{n+1}| \frac{\partial \H_n^{c}}{\partial u_n} \U_n\ut{ST_2} \frac{\partial \H_n^{c}}{\partial u_n}|\psi_n}\bigg)  \delta_{n,m} \dt \bigg\}.
\end{align}
The Hessian expression for $\U_n\ut{ST_2}$ is thus much more cumbersome than that for $\U_n\ut{ST_1}$.

\subsection{Derivatives for Regularizations}
In many applications, it is advantageous to regularize either or both the control amplitude and its temporal derivative. This requires additional terms in the cost functional objective, imposition of discretization, and calculation of the respective derivatives. As with the propagator, the chosen form of the implementation scheme changes the derivative calculations.

The amplitude regularization is straightforward,
\begin{align}
J_\alpha &= \frac{\alpha}{2} \int_{0}^{T} u(t)^2 \d t \rightarrow \frac{\alpha}{2} \dt \sum_{i=1}^{\nt} u_i^2, \\
\frac{\partial J_\alpha}{\partial u_n} &= \alpha\dt u_n, \quad \quad
\frac{\partial^2 J_\alpha}{\partial u_m \partial u_n} = \alpha\dt \delta_{n,m},
\end{align}
where $\alpha$ is a weighting factor.
The derivative regularization is a bit more involved because of the end points
\begin{align}
&J_\gamma = \frac{\gamma}{2} \int_{0}^{T} \dot{u}(t)^2 \d t 
 \rightarrow \frac{\gamma}{8\dt}\bigg(\sum_{i=2}^{\nt - 1}(u_{i+1} - u_{i-1})^2 \notag \\
&\left[-3 u_1 +4 u_2 -u_3\right]^2 + \left[3 u_{\nt} - 4 u_{\nt-1} +u_{\nt-2}\right]^2 \bigg),
\end{align}
where we used forward (backward) difference approximations for the first (last) point and center approximations for the bulk, all to order $\mathcal{O}(\dt^2)$.
The derivative with respect to the first and last three indices is different from the bulk. 
The resulting gradient written in vector form is
\begin{align}
\nabla J_\gamma = \frac{\gamma}{4\dt}
 \begin{bmatrix}
	10u_1 - 12 u_2 + 2u_3 \\
	-12 u_1 + 17 u_2 - 4 u_3 - u_4 \\
	2 u_1 - 4 u_2 + 3 u_3 - u_5 \\
	\vdots \\
	2 u_n - u_{n-2} + u_{n+2} \\
	\vdots\\
	2 u_{\nt} - 4u_{\nt-1} + 3 u_{\nt -2} - u_{\nt-4} \\
	-12 u_{\nt} + 17 u_{\nt -1}  - 4 u_{\nt-2} - u_{\nt - 3}   \\
	10 u_{\nt} - 12 u_{\nt-1} + 2 u_{\nt-2}
\end{bmatrix},
\end{align}
where the vertical dots extend over the bulk points.
Similarly the three first and last Hessian rows are different from the bulk. In a stacking notation where the indices denote the rows, we obtain
\begin{align}
\nabla^2 J_\gamma &=
\begin{bmatrix}
[\nabla^2 J_\gamma]_{1:3} \\
\hline
[\nabla^2 J_\gamma]_{4:\nt-3} \\
\hline
[\nabla^2 J_\gamma]_{\nt-2:\nt}, \\
\end{bmatrix}
\end{align}
where the matrices evaluate to
\begin{align}
[\nabla^2 J_\gamma]_{1:3} & = \frac{\gamma}{4\dt}  
\begin{bmatrix}
10 & -12 & 2 & 0 & 0 & 0 & \dots\\
-12 & 17 & -4 & -1 & 0 &0 &  \dots\\
2 & -4 & 3 & 0 & -1 & 0 &  \dots
\end{bmatrix}, \\
[\nabla^2 J_\gamma]_{4:\nt-3}  &=  \notag\\
&\frac{\gamma}{4\dt}  
\begin{bmatrix}
0 & -1 & 0 & 2 & 0 & -1 & 0 & 0 &\dots \\
  &  & \ddots  && \ddots & &\ddots & & \\
\dots &0 & 0 & -1 & 0 & 2 & 0 & -1 & 0  \\
\end{bmatrix}, \\
[\nabla^2 J_\gamma]_{\nt-2:\nt} & = \frac{\gamma}{4\dt}  
\begin{bmatrix}
\dots &  0 & -1 & 0 & 3 & -4 & 2 \\
\dots &  0 & 0& -1 & -4 & 17 & -12 \\
\dots &  0 & 0& 0 & 2 & -12 & 10 \\
\end{bmatrix}, 
\end{align}
and the dots denote continuation of the number they point to.

\bibliography{/Users/au446513/projects/bibmaster/references.bib}

\end{document}